\documentclass[aps,prb,showpacs,twocolumn]{revtex4-1}
\usepackage{amsmath,amssymb,bm,dcolumn,epsfig,graphicx,longtable}
\begin{document}
\title{Approximations to the exact exchange potential: KLI versus semilocal}
\author{Fabien Tran}
\author{Peter Blaha}
\affiliation{Institute of Materials Chemistry, Vienna University of Technology,
Getreidemarkt 9/165-TC, A-1060 Vienna, Austria}
\author{Markus Betzinger}
\author{Stefan Bl\"{u}gel}
\affiliation{Peter-Gr\"{u}nberg Institut and Institute for Advanced Simulation,
Forschungszentrum J\"{u}lich and JARA, D-52425 J\"{u}lich, Germany}

\begin{abstract}

In the search for an accurate and computationally efficient approximation to
the exact exchange potential of Kohn-Sham density functional theory,
we recently compared various semilocal exchange potentials to the exact one
[F. Tran \textit{et al}., Phys. Rev. B \textbf{91}, 165121 (2015)].
It was concluded that the Becke-Johnson (BJ) potential is a very good starting
point, but requires the use of empirical parameters to obtain good agreement
with the exact exchange potential. In the present work, we extend the comparison
by considering the Krieger-Li-Iafrate (KLI) approximation,
which is a beyond-semilocal approximation. It is shown that overall the KLI and
BJ-based potentials are the most reliable approximations to the exact exchange
potential, however, sizeable differences, especially for the antiferromagnetic
transition-metal oxides, can be obtained.

\end{abstract}

\pacs{71.15.Ap, 71.15.Mb, 71.20.-b}
\maketitle

\section{\label{introduction}Introduction}

Due to its rather low cost/accuracy ratio, the
Kohn-Sham (KS) version of density functional theory\cite{HohenbergPR64,KohnPR65}
is the most widely used method for the calculation of the geometrical and
electronic properties of matter nowadays. The reliability of the results of a KS
calculation depends mainly on the chosen approximation for the
exchange-correlation (xc) energy $E_{\text{xc}}$ and potential
$v_{\text{xc},\sigma}$ ($\sigma$ is the spin index). The geometric properties
are mostly (but not exclusively\cite{KimPRL13}) determined by the energy
$E_{\text{xc}}$, while the electronic structure
is governed by the potential $v_{\text{xc},\sigma}$.\cite{KuemmelRMP08,CohenCR12}

In the KS method, the xc potential is multiplicative since it is calculated
as the functional derivative of the xc functional with respect to the electron
density $\rho_{\sigma}$ ($v_{\text{xc},\sigma}=\delta E_{\text{xc}}/\delta\rho_{\sigma}$).
From the variational point of view this is more restrictive than taking the
derivative with respect to the orbitals $\psi_{i}^{\sigma}$
($\hat{v}_{\text{xc}}\psi_{i}^{\sigma}=\delta E_{\text{xc}}/\delta\psi_{i}^{\sigma*}$),
like in the generalized KS framework,\cite{SeidlPRB96} which leads to non-multiplicative
xc potentials in the case of implicit functionals of the electron density.
A straightforward analytical calculation of
$v_{\text{xc},\sigma}=\delta E_{\text{xc}}/\delta\rho_{\sigma}$ is possible for
explicit functionals of $\rho_{\sigma}$ like those of the local density
approximation (LDA) or generalized gradient approximation (GGA). However, for
implicit functionals of $\rho_{\sigma}$, like meta-GGA (MGGA) or the Hartree-Fock (HF)
exchange [which is also the exact exchange (EXX) in the KS theory], such a
direct analytical calculation of the xc potential is not possible and one has
to resort to the optimized effective method\cite{SharpPR53} (OEP) which
consists of solving integro-differential equations to get $v_{\text{xc},\sigma}$.

Since the EXX energy in the KS method is known:
\begin{widetext}
\begin{equation}
E_{\text{x}}^{\text{EXX}} =
-\frac{1}{2}\sum_{\sigma}\sum_{i=1}^{N_{\sigma}}\sum_{j=1}^{N_{\sigma}}\int\int
\frac{
\psi_{i}^{\sigma*}(\bm{r})\psi_{j}^{\sigma}(\bm{r})
\psi_{j}^{\sigma*}(\bm{r}')\psi_{i}^{\sigma}(\bm{r}')}
{\left\vert\bm{r}-\bm{r}'\right\vert}d^{3}rd^{3}r',
\label{ExEXX}
\end{equation}
\end{widetext}
the OEP applied to EXX gives us access
to the exact KS exchange potential (thereafter called EXX-OEP),
and implementations have been reported for
molecules and periodic systems (see Refs.~\onlinecite{KuemmelRMP08,Engel} for
reviews and, e.g., Refs.~\onlinecite{BetzingerPRB11,GidopoulosPRA12,EngelJCP14}
for recent implementations).

Since the implementation of a numerically stable OEP approach is quite involved
(see, e.g., Ref. \onlinecite{BetzingerPRB11})
and since an EXX-OEP calculation formally scales with the fourth power
of the system size, an accurate, reliable, and fast approximation to EXX-OEP
is of high interest.
In a recent study,\cite{TranPRB15} we showed that among various semilocal
approximations for the exchange potential, the best agreement with EXX-OEP in
solids was obtained with a modification of the potential
proposed by Becke and Johnson\cite{BeckeJCP06} (BJ). The conclusions were based
on a comparison of the total energy, electronic structure, magnetic moment, and
electric-field gradient (EFG) for a set of six solids.

In this work, we proceed by a comparison of the EXX-OEP with an approximate
form suggested by Krieger, Li, and Iafrate
\cite{KriegerPLA90a,KriegerPRA92a,KriegerPRA92b,LiPRA93} (KLI).
The KLI approximation to OEP, which has also been used for functionals other
than EXX (e.g., self-interaction corrected
\cite{LiPRB91,TongPRA97,GarzaJCP00,PatchkovskiiJCP01}
or MGGA\cite{ArbuznikovCPL03,EichJCP14,YangPRB16} functionals) is an
interesting alternative to the OEP since it avoids the numerical difficulties
of EXX-OEP (very recent works are Refs.
\onlinecite{FuksPRA11,ArnoldJPCM11,QianPRB12,VilhenaPRA12,SchmidtPCCP14,KraislerJCP15,KimBKCS15,KimPCCP15,SchmidtPRB16}
in the case of EXX and Refs.~\onlinecite{EichJCP14,YangPRB16} for MGGAs).
However, comparisons between the EXX-OEP and the KLI approximation
(EXX-KLI in the following) concern mainly atoms and light molecules/clusters
\cite{KriegerPRA92a,KriegerPRA92b,LiPRA93,GritsenkoPRA95,GraboCPL95,GraboME97,EngelPRA00,DellaSalaJCP01,GruningJCP02,KuemmelPRL03,KuemmelPRB03,MakmalPRB09,Engel,RyabinkinPRL13,KohutJCP14}
and only a few such comparisons were done for periodic systems.
\cite{EngelPRL09,Engel,EngelJCP14,RigamontiPRB15}
From most of these studies, it was concluded that EXX-KLI is a good
approximation to EXX-OEP, however, in Refs.~\onlinecite{EngelPRL09,Engel}
Engel pointed out that in bulk Si and FeO the EXX-KLI potential can not fully
reproduce the aspherical features around the atoms seen in the EXX-OEP.
Overall, the number and variety of systems used in these comparisons between
EXX-OEP and EXX-KLI is not very exhaustive, and since the EXX-KLI approximation
is easier to implement and computationally more advantageous
than EXX-OEP, a more systematic comparison between these two potentials giving a
better idea of the accuracy of EXX-KLI would be certainly useful.

To this end, the EXX-KLI potential has been implemented in an all-electron code
for solid-state calculations and applied, along with the EXX-OEP, to various
types of solids. In addition, we compare the EXX-KLI to the semilocal
potentials already analyzed in our previous work, and pursue the question
which of these potentials is the best approximation to EXX-OEP.
This is an important question since the semilocal potentials are
computationally much faster than EXX-KLI.

The paper is organized as follows. Section~\ref{theory} provides a short
description of the potentials as well as the computational details.
Then, the results are presented and discussed in Sec.~\ref{results},
while Sec.~\ref{summary} gives the summary.

\section{\label{theory}Theory and computational details}

The functional derivative of an only implicit functional of the
density with respect to the density can be obtained by making use of
the OEP approach.\cite{KuemmelRMP08,Engel}
It leads to a complicated integro-differential equation, which involves
response functions for the KS orbitals and density.
The KLI approximation to the OEP equation consists of replacing all
orbital energies differences $\varepsilon_{j\sigma}-\varepsilon_{i\sigma}$
in the response function by the same constant
$\Delta\varepsilon_{\sigma}$.\cite{KriegerPLA90a,KriegerPRA92a,KriegerPRA92b}
In the case of EXX, or also MGGA functionals,
\cite{ArbuznikovCPL03,EichJCP14,YangPRB16} the equations become much
more simple to solve since the sum over the (infinite) number of unoccupied
states can be collapsed, so that the need of unoccupied states can be
completely avoided. The KLI equations for EXX are
\begin{eqnarray}
v_{\text{x},\sigma}^{\text{EXX-KLI}}(\bm{r}) & = &
v_{\text{x},\sigma}^{\text{S}}(\bm{r}) +
\frac{1}{\rho_{\sigma}(\bm{r})}\sum_{i=1}^{N_{\sigma}}
\left\vert\psi_{i}^{\sigma}(\bm{r})\right\vert^{2} \nonumber \\
& & \times\left(\langle\psi_{i}^{\sigma}\vert v_{\text{x},\sigma}^{\text{EXX-KLI}}\vert
\psi_{i}^{\sigma}\rangle - \langle\psi_{i}^{\sigma}
\vert\hat{v}_{\text{x},\sigma}^{\text{HF}}\vert
\psi_{i}^{\sigma}\rangle\right), \nonumber \\
\label{vxKLI1}
\end{eqnarray}
where $v_{\text{x},\sigma}^{\text{S}}$ is
the Slater potential\cite{SlaterPR51}
\begin{eqnarray}
v_{\text{x},\sigma}^{\text{S}}(\bm{r}) & = &
-\frac{1}{\rho_{\sigma}(\bm{r})}
\sum_{i=1}^{N_{\sigma}}\sum_{j=1}^{N_{\sigma}}
\psi_{i}^{\sigma*}(\bm{r})\psi_{j}^{\sigma}(\bm{r}) \nonumber\\
& & \times
\int\frac{\psi_{j}^{\sigma*}(\bm{r}')\psi_{i}^{\sigma}(\bm{r}')}
{\left\vert\bm{r}-\bm{r}'\right\vert}d^{3}r'.
\label{vxS}
\end{eqnarray}
The sum in the second term of Eq.~(\ref{vxKLI1}) should
in principle run over all occupied orbitals, however in order to ensure the
correct asymptotic behavior of the potential far from the nuclei it has been
rather common for molecular calculations to discard the highest occupied
orbital from this sum.\cite{KriegerPLA90a} This is what has also been done
for the calculations on periodic solids reported in
Refs.~\onlinecite{BylanderPRL95,BylanderPRB95,BylanderPRB96,BylanderPRB97},
however it is obvious that in this case this does not make sense,
since removing the highest occupied orbital at one $\bm{k}$-point (or set of
equivalent $\bm{k}$-points) would have no effect in the limit of a dense
$\bm{k}$-mesh. Therefore, we chose to include all occupied orbitals in
the sum in Eq.~(\ref{vxKLI1}) for the present work.

We mention that the potential known as localized HF (LHF,
Ref.~\onlinecite{DellaSalaJCP01}), or alternatively as the common energy
denominator approximation (CEDA, Ref.~\onlinecite{GritsenkoPRA01}), has the
same form as Eq.~(\ref{vxKLI1}), the difference being that the second term
consists of a double sum over the orbitals instead of only one,
therefore, the EXX-KLI potential can also be considered as a simplification of
the LHF/CEDA potential. Other alternative derivations of Eq.~(\ref{vxKLI1}) can
be found in Refs.~\onlinecite{KriegerPRA92a,BylanderPRL95,NagyPRA97}.

A certain number of studies about EXX-KLI have been
published in the literature, but among them only a few concerned periodic
systems. These works on periodic systems are now summarized. Plane-wave
pseudopotential calculations were reported by Bylander and Kleinman
\cite{BylanderPRL95,BylanderPRB95,BylanderPRB96,BylanderPRB97}
on the semiconductors Si, Ge, and GaAs, more recently by
Engel and co-workers\cite{EngelPRB01,EngelPRL09,Engel,EngelJCP14,EngelPRB14}
on Al, Si, FeO, and slab systems, as well as by Natan\cite{NatanPCCP15} on C, Si,
and polyacetylene. S\"{u}le \textit{et al}.\cite{SuleJCP00} applied
EXX-KLI to polyethylene using a code based on Gaussian basis functions.
Fukazawa and Akai\cite{FukazawaJPCM10,FukazawaJPCM15} reported KLI
results for alkali and magnetic metals (Li, Na, K, Fe, Co, and Ni) and
antiferromagnetic MnO
which were obtained with a code based on the Korringa-Kohn-Rostoker Green
function method, while the details of a EXX-KLI implementation within the
projected-augmented-wave formalism are available in the work of Xu and
Holzwarth.\cite{XuPRB11}

For the purpose of the present work, the EXX-KLI potential,
Eq.~(\ref{vxKLI1}), has been implemented
into the all-electron code WIEN2k,\cite{WIEN2k} which is based on the
linearized augmented plane-wave (LAPW) method.\cite{AndersenPRB75,Singh,Blugel}
The implementation of the Slater potential [Eq.~(\ref{vxS})] into the WIEN2k
code has been reported recently\cite{TranJCTC15} and the same techniques
were used for the additional term in Eq.~(\ref{vxKLI1}). Details of the
equations specific for the LAPW basis set can be found in
the Supplemental Material.\cite{SM_KLI}
Here, we just mention that the implementation of Eq.~(\ref{vxKLI1}) is exact
and is based on the pseudocharge method\cite{WeinertJMP81,MassiddaPRB93}
combined with the technique proposed in Refs.~\onlinecite{OnidaPRL95,SpencerPRB08}
to treat the Coulomb singularity in the integrals involving the HF operator
(see also Ref.~\onlinecite{TranPRB11}).
As done by S\"{u}le \textit{et al}.\cite{SuleJCP00} and Engel,\cite{Engel16}
the self-consistent-field (SCF) procedure to solve the KS equations with the
EXX-KLI potential $v_{\text{x},\sigma}^{\text{EXX-KLI}}$ was done by using
$v_{\text{x},\sigma}^{\text{EXX-KLI}}$ from the previous iteration to calculate the
integrals $\langle\psi_{i}^{\sigma}\vert v_{\text{x},\sigma}^{\text{EXX-KLI}}\vert
\psi_{i}^{\sigma}\rangle$ on the right-hand side of Eq.~(\ref{vxKLI1}).
(Another possibility would have been to solve a set of linear
equations at each iteration.\cite{KriegerPLA90a})
We also mention that the SCF convergence could
be achieved much more efficiently by using an inner/outer loops procedure
similar to the one described in Ref.~\onlinecite{BetzingerPRB10} for the HF
method.

The EXX-OEP calculations, which will serve as reference for the discussion
of the results, were done with the
FLEUR code\cite{FLEUR} that is also based on the LAPW method.
The implementation of the EXX-OEP method in FLEUR employs an auxiliary basis,
the mixed product basis, for representing the EXX-OEP, and as shown in
Refs.~\onlinecite{BetzingerPRB11,BetzingerPRB12,BetzingerPRB13,FriedrichPRA13},
very well converged all-electron EXX-OEP could be obtained thanks to an accurate
and efficient construction of the KS orbitals and density response.

The semilocal calculations were done with the following exchange-only potentials
$v_{\text{x},\sigma}$. The LDA potential,\cite{KohnPR65} which is exact for the
homogeneous electron gas, depends only on $\rho_{\sigma}$. The potentials of
Perdew, Burke, and Ernzerhof\cite{PerdewPRL96} (PBE), Engel and
Vosko\cite{EngelPRB93} (EV93), and Armiento and K\"{u}mmel\cite{ArmientoPRL13}
(AK13) are functional derivatives of functionals $E_{\text{x}}$ of the GGA form
and hence depend on $\rho_{\sigma}$ and its first two derivatives
$\nabla\rho_{\sigma}$ and $\nabla^{2}\rho_{\sigma}$.
In Ref.~\onlinecite{TranPRB15}, a generalization of the BJ
potential\cite{BeckeJCP06} (gBJ) was proposed as an approximation to the EXX-OEP
in solids. The gBJ potential, which is of the MGGA form since it depends on the
kinetic-energy density $t_{\sigma} = \left(1/2\right)\sum_{i=1}^{N_{\sigma}}
\nabla\psi_{i}^{\sigma*}\cdot\nabla\psi_{i}^{\sigma}$, was shown to be more
accurate than the GGA potentials mentioned just above (the test set of solids was
composed of C, Si, BN, MgO, Cu$_{2}$O, and NiO). However, this good
agreement with EXX-OEP was achieved by tuning the three empirical parameters
($\gamma$, $c$, and $p$) in gBJ, and it was shown that a set of parameters that
is good for a property or group of solids may not give good results for other
properties/solids. For instance, a good agreement with EXX-OEP for the magnetic
moment in NiO requires values for $(\gamma,c,p)$ that are very different from
those for the band gap or total energy.\cite{TranPRB15} Furthermore, it was
also shown that meaningful results for the band gap and EFG in Cu$_{2}$O
could only be obtained by considering the universal
correction to the gBJ potential.\cite{RasanenJCP10} For the present work, we
decided to consider only one of the four parameterizations of the gBJ potential
discussed in Ref.~\onlinecite{TranPRB15}, namely, the one for the total
energy [$(\gamma,c,p)=(0.6,1.0,0.60)$].
Showing also the results obtained with the
parameterization that is on average slightly more accurate for the band gap
[$(\gamma,c,p)=(1.4,1.1,0.50)$] would not change the conclusions of the present
work. The two other sets of parameters were proposed for NiO and
Cu$_{2}$O specifically and lead to very bad results for other systems
such that they are of limited interest.

The convergence parameters of the calculations with WIEN2k and FLEUR, like the
size of the basis set or the number of $\bm{k}$-points for the integrations
in the Brillouin zone, were chosen such that the results are well converged
(e.g., within $\sim0.03$~eV for the band gap).
The solids of the test set are listed in Table~S1 of the Supplemental Material,
\cite{SM_KLI} along with their space group and geometrical parameters.
The core electrons (also indicated in Table~S1) were treated fully
relativistically (i.e., including spin-orbit coupling), while a
scalar-relativistic treatment\cite{KoellingJPC77} was used for the valence electrons.

\section{\label{results}Results and discussion}

\subsection{\label{EXXtotalenergy}EXX total energy and electron density}

\begin{table}
\caption{\label{table_statistics}Average over the solids of the errors
(with respect to EXX-OEP) in the EXX total energy, electron density,
KS fundamental band gap, and energy position of the core states.
See text for details.}
\begin{ruledtabular}
\begin{tabular}{lcccccc}
           & EXX-KLI & LDA & PBE & EV93 & AK13 & gBJ \\
\hline
EXX total energy \\
ME (mRy/cell)  & 18 & 218 & 139 & 87 & 84 & 55 \\
MAE (mRy/cell) & 19 & 218 & 139 & 87 & 84 & 55 \\
\hline
Electron density \\
ME & 0.9 & 3.1 & 2.1 & 1.6 & 1.8 & 0.9 \\
\hline
Band gap \\
ME (eV)    & -0.58 & -1.84 & -1.36 & -0.96 & 0.39 & -0.63 \\
MAE (eV)   & 0.58 & 1.84 & 1.36 & 1.03 & 1.20 & 0.71 \\
\hline
Core states \\
MMRE (\%)  & 0.2 & 1.1 & 0.3 & 0.0 & -0.6 & -0.3 \\
MMARE (\%) & 0.6 & 1.3 & 0.5 & 0.6 & 0.9 & 0.5 \\
\end{tabular}
\end{ruledtabular}
\end{table}

\begin{figure}[t]
\includegraphics[width=\columnwidth]{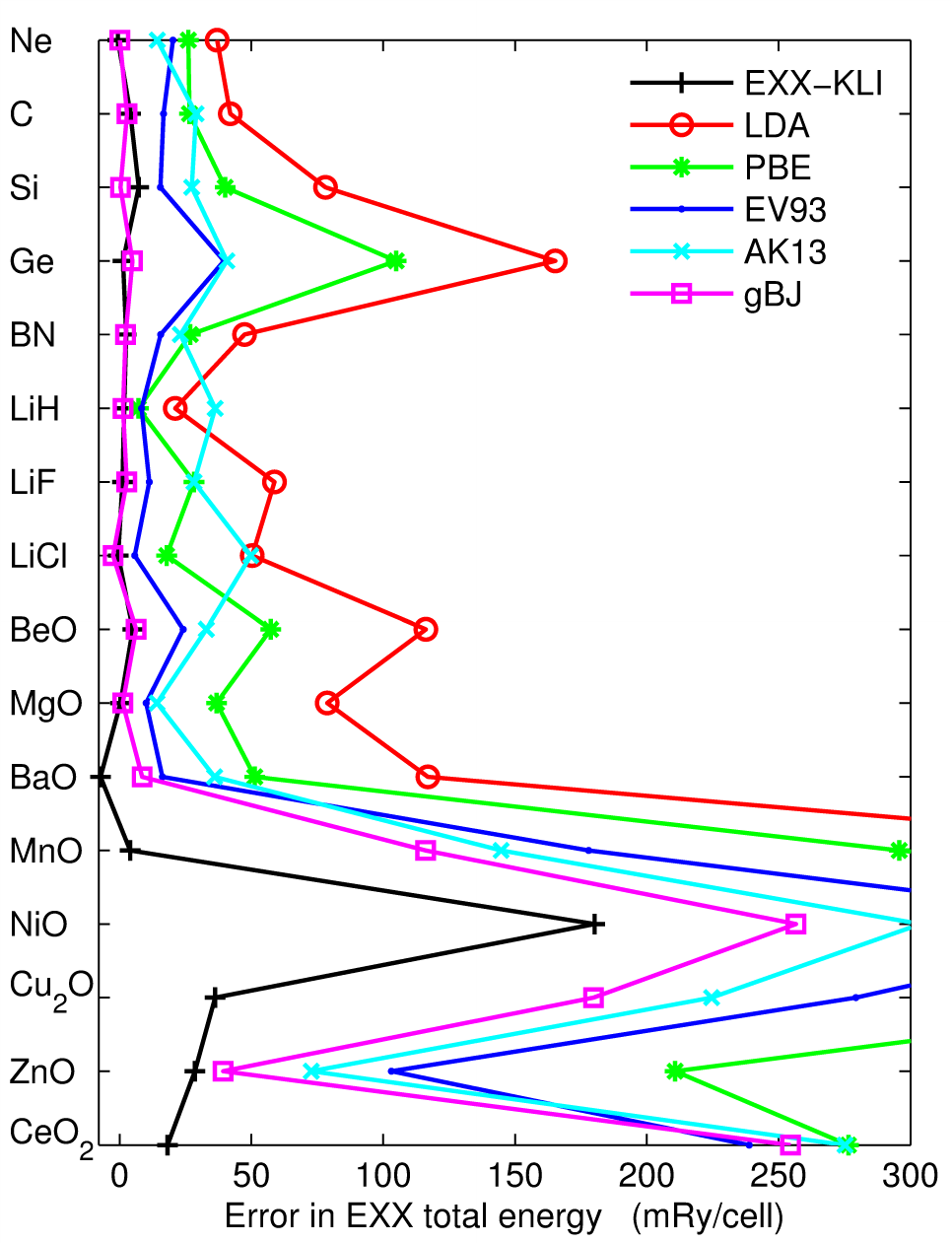}
\caption{\label{fig_Etot}Error (in mRy/cell) in the EXX total energy
calculated with orbitals generated from approximate exchange potentials with
respect to the values obtained with the EXX-OEP orbitals.}
\end{figure}
\begin{figure}[t]
\includegraphics[width=\columnwidth]{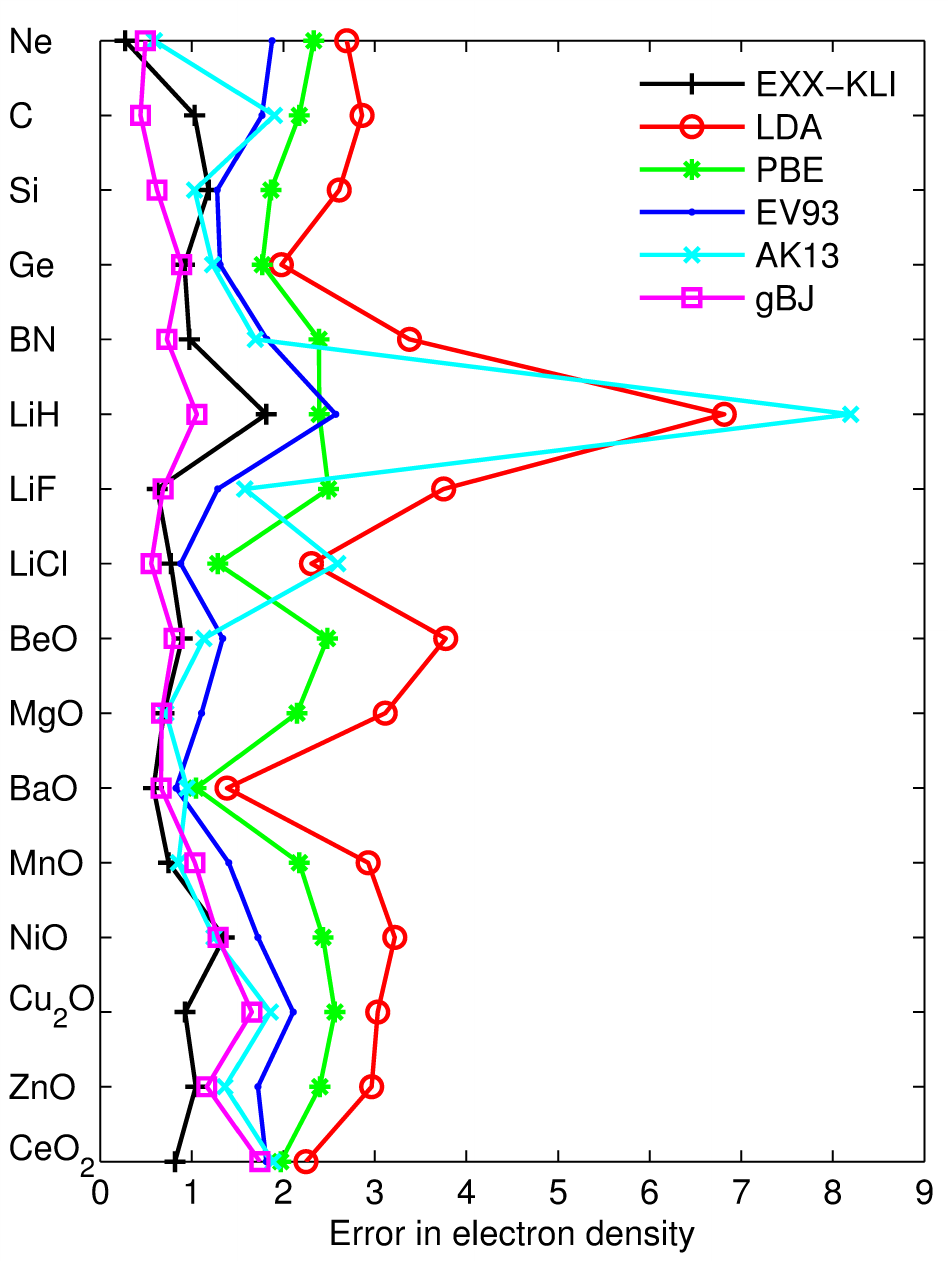}
\caption{\label{fig_rho1rho2int1}Integrated density difference
as defined by Eq.~(\ref{rhodiff}).}
\end{figure}

We begin the discussion of the results with the EXX total energy
$E_{\text{tot}}^{\text{EXX}}$. The results are shown graphically in
Fig.~\ref{fig_Etot} for each solid (see Table~S2 of the Supplemental
Material\cite{SM_KLI} for the numerical values) and Table~\ref{table_statistics}
contains the mean error (ME) and mean absolute error (MAE).
As in Refs.~\onlinecite{TranPRB15,TranJCTC15}, the EXX total energy expression
[Eq.~(\ref{ExEXX}) for $E_{\text{x}}$ and no correlation]
has been evaluated with the orbitals generated from various potentials.
The error is with respect to the value obtained with the EXX-OEP orbitals:
$E_{\text{tot}}^{\text{EXX}}[\{\psi_{i}^{\text{approx}}\}]-
E_{\text{tot}}^{\text{EXX}}[\{\psi_{i}^{\text{EXX-OEP}}\}]$,
where $E_{\text{tot}}^{\text{EXX}}[\{\psi_{i}^{\text{EXX-OEP}}\}]$ is the EXX
total energy calculated with the EXX-OEP orbitals and
$E_{\text{tot}}^{\text{EXX}}[\{\psi_{i}^{\text{approx}}\}]$ is the value
obtained with orbitals obtained by one of the approximate exchange potentials.
From the results we can see that the smallest errors with respect to EXX-OEP
are obtained with the EXX-KLI and gBJ orbitals. With the exception of NiO, EXX-KLI
leads to errors which are below 50 mRy/cell, and the MAE is about
20~mRy/cell. gBJ leads to very similar errors except for the
transition-metal oxides and CeO$_{2}$ for which the errors are clearly larger (up to
$\sim260$~mRy/cell for NiO and CeO$_{2}$). These differences between the EXX-KLI and gBJ total
energies for the transition-metal oxides are in line with the results for the
electronic structure which show that EXX-KLI is much more accurate than gBJ
(see below). The MAE with the gBJ potential of 55~mRy/cell is three times larger
than for EXX-KLI. The orbitals generated by the other potentials
lead to EXX total energies that are much higher (i.e., less negative)
and to MAE of 218 (LDA), 139 (PBE), 87 (EV93), and 84 (AK13) mRy/cell.
As a technical note, we remark that a few of the errors in Fig.~\ref{fig_Etot}
(Table~S2) obtained with EXX-KLI and gBJ are slightly negative.
In principle this should not occur, since among all sets of orbitals generated
by a multiplicative potential, the EXX-OEP orbitals should, by definition, lead
to the most negative EXX total energy. These negative values, which are anyway
tiny and of no importance for the discussion, might be due to
some minor (but unavoidable) incompatibilities between the two LAPW codes,
e.g., details of the basis set or the integration methods.

Using the EXX total energy is a way to quantify with a single number the
difference in shape between two sets of orbitals. An alternative is to
consider the difference between the electron densities as follows:
\begin{equation}
\frac{100}{N}\int\limits_{\Omega}
\left\vert\rho^{\text{approx}}(\bm{r})-\rho^{\text{EXX-OEP}}(\bm{r})\right\vert d^{3}r,
\label{rhodiff}
\end{equation}
where $N=\int_{\Omega}\rho d^{3}r$ is the number of electrons in the unit cell
$\Omega$ and the multiplication by 100 makes the numerical values more convenient.
The absolute value of the integrand is taken in order to avoid
cancellation between positive and negative values of
$\rho^{\text{approx}}-\rho^{\text{EXX-OEP}}$.
The results of Eq.~(\ref{rhodiff}) for the different approximate potentials
and solids are displayed in Fig.~\ref{fig_rho1rho2int1}, while
The ME over the solids is shown in Table~\ref{table_statistics}.
The main observation is the same as with the EXX total energy, namely,
the EXX-KLI and gBJ potentials lead to the smallest errors on average.
However, both potentials lead to the same ME (0.9), which was not the case for the
EXX total energy; one of the reasons is that Eq.~(\ref{rhodiff})
is normalized with the number of electrons that is much larger for the
transition-metal oxides and CeO$_{2}$, such that the large spreads in the
errors observed in Fig.~\ref{fig_Etot} become similar to the other solids.
This is again with LDA
that the largest ME (3.1) is obtained. From Fig.~\ref{fig_rho1rho2int1} we can see that
the LDA and AK13 potentials lead to very large density difference for LiH, which
should mainly be due to the Li-$1s$ core states (see Sec.~\ref{electronicproperties}).

Thus, we can conclude that in terms of EXX total energy and integrated electron
density difference, the EXX-KLI and gBJ potentials are on average the closest
to the EXX-OEP.

\subsection{\label{electronicproperties}Electronic properties}

\begin{figure}[t]
\includegraphics[width=\columnwidth]{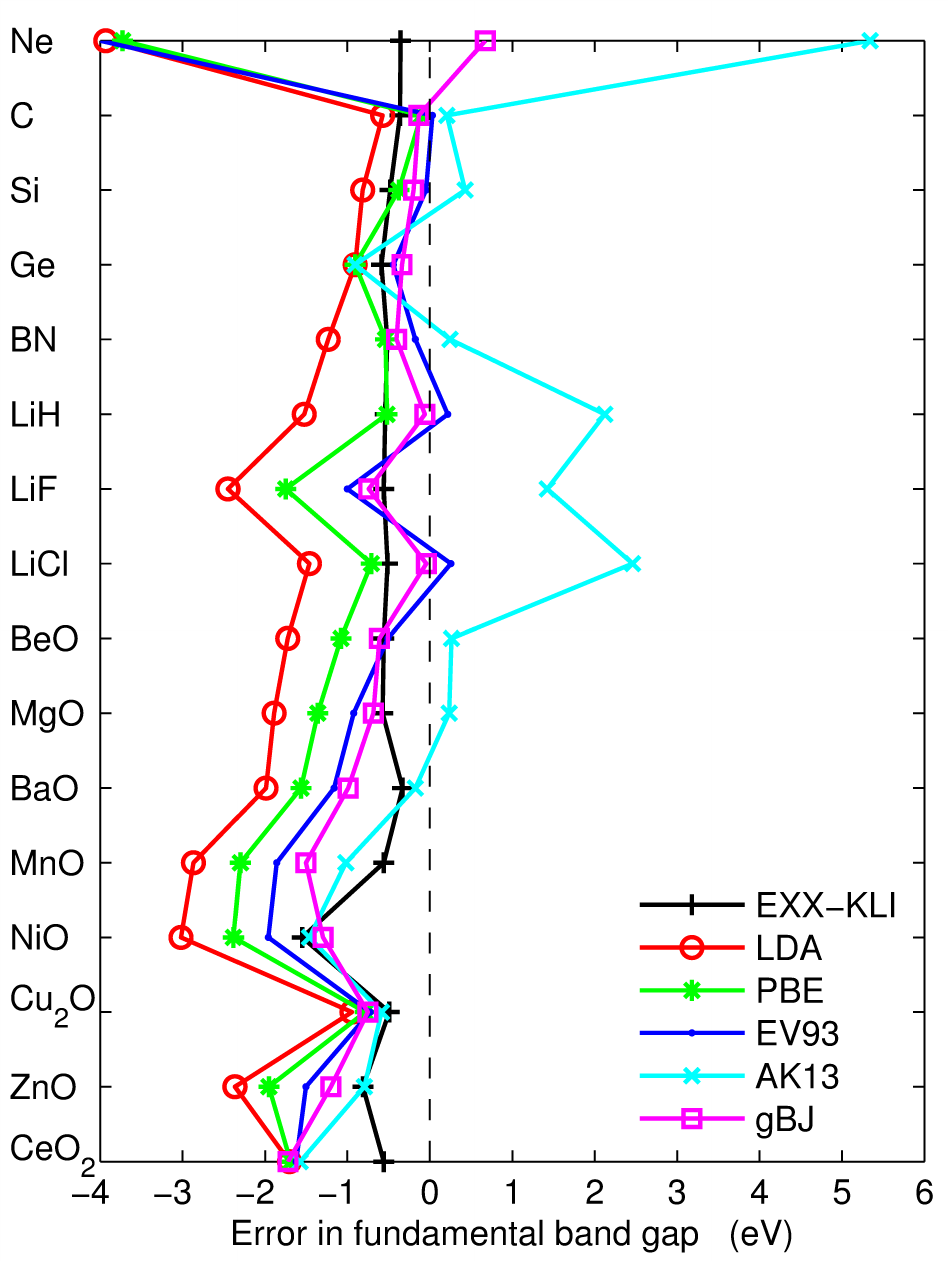}
\caption{\label{fig_gap}Error (in eV) in the KS fundamental band gap calculated
with approximate exchange potentials with respect to the EXX-OEP values.}
\end{figure}

Turning now to the electronic band structure, the results for the
KS fundamental band gap, defined as the conduction band minimum minus the
valence band maximum, are shown in Fig.~\ref{fig_gap} and Table~S3 of the
Supplemental Material.\cite{SM_KLI} The LDA and standard GGAs like PBE are
known to underestimate the band gap by a rather large amount in solids compared to
EXX-OEP.\cite{StaedelePRL97,BetzingerPRB11,HollinsPRB12,TranPRB15,EngelIJQC16}
Such an underestimation is indeed observed for all solids considered in the
present work, and it is the largest, between 2 and 4~eV, for Ne, LiF, MnO,
NiO, and ZnO. The GGA EV93 exchange functional,\cite{EngelPRB93} which was
designed to have a functional derivative which resembles the EXX-OEP in atoms,
increases the band gap with respect to the LDA and standard GGAs potentials
such that a better agreement with EXX-OEP is usually obtained
(see Fig.~\ref{fig_gap} and, e.g.,
Refs.~\onlinecite{DufekPRB94,TranJPCM07,TranPRB15,TranJCTC15}).
An exception is Ne since the EV93 band gap is slightly smaller than
the LDA and PBE band gaps. In Table~\ref{table_statistics}, the ME
and MAE for the band gap are reduced for EV93
compared to LDA and PBE, but there is still a non-negligible
underestimation of $-0.96$~eV on average. The AK13 potential
also improves over LDA and PBE on average (ME and MAE of 0.39 and
1.20~eV, respectively), but leads to rather important overestimations for
Ne, LiH, LiF, and LiCl, that are due to the excessively large positive values
of the AK13 potential in the interstitial region
as discussed in Refs.~\onlinecite{TranPRB15,TranJCTC15}.

The smallest MAE in Table~\ref{table_statistics} for the band gap are obtained
with the EXX-KLI and gBJ potentials, which lead to values in the range 0.6-0.7~eV,
while the other potentials lead to MAE above 1~eV.
Also, the error for Ne is strongly reduced compared
to the other methods (see Fig.~\ref{fig_gap}). However, by looking at the
detailed results, we can see that there are some noticeable differences in
the trends in the EXX-KLI and gBJ band gaps. In particular, the curve of the
error for gBJ has a similar shape as for LDA, PBE, and EV93 in the sense that
the error clearly varies from one solid to the other, while this is not the
case with EXX-KLI since the error is in a narrow window around
$-0.5$~eV for most solids except NiO ($-1.5$~eV).
This is a quite interesting observation since the error in the band gap with
EXX-KLI seems to be more predictable than with the other potentials.
Direct comparisons between EXX-OEP and EXX-KLI were also reported by Engel
and co-workers.\cite{MakmalPRB09,EngelPRL09}
In Ref.~\onlinecite{MakmalPRB09}, the EXX-KLI gap was reported to be
too small in the CO and BeO molecules by 0.47 and 0.24~eV, respectively,
while in Ref.~\onlinecite{EngelPRL09} a metallic ground-state for
antiferromagnetic FeO was obtained with EXX-KLI, which is a qualitatively
wrong result since EXX-OEP (with LDA correlation added) leads to a band
gap of 1.66 eV.\cite{EngelPRL09} Actually, we could confirm (with our
implementation) that EXX-KLI leads to no band gap in FeO, which means that
in this respect, semilocal potentials can perform better since gBJ
(with 0.62~eV) and some others\cite{TranJCTC15} open a band gap.
We also mention that for CoO, we obtained a EXX-KLI band gap of 0.48 eV,
which is about 2~eV smaller than the EXX-OEP value reported by
Engel,\cite{EngelPRL09} while AK13 and gBJ lead to band gaps of 1.37 and
1.18~eV, respectively.

\begin{figure}[t]
\includegraphics[width=\columnwidth]{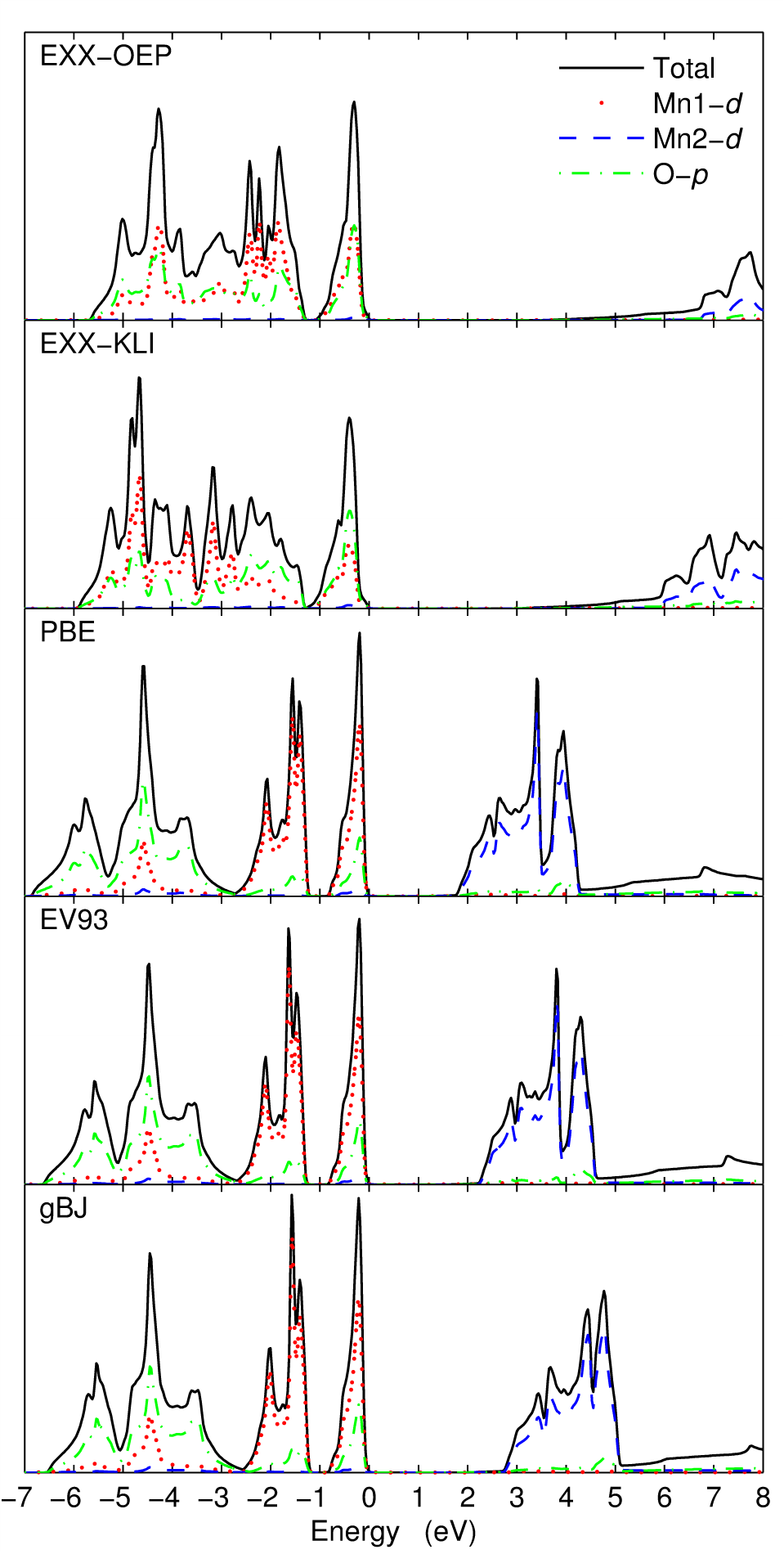}
\caption{\label{fig_MnO_DOS}Spin-up DOS of MnO.
Mn1 is the Mn atom with majority spin-up electrons.
The Fermi energy is set at zero.}
\end{figure}
\begin{figure}[t]
\includegraphics[width=\columnwidth]{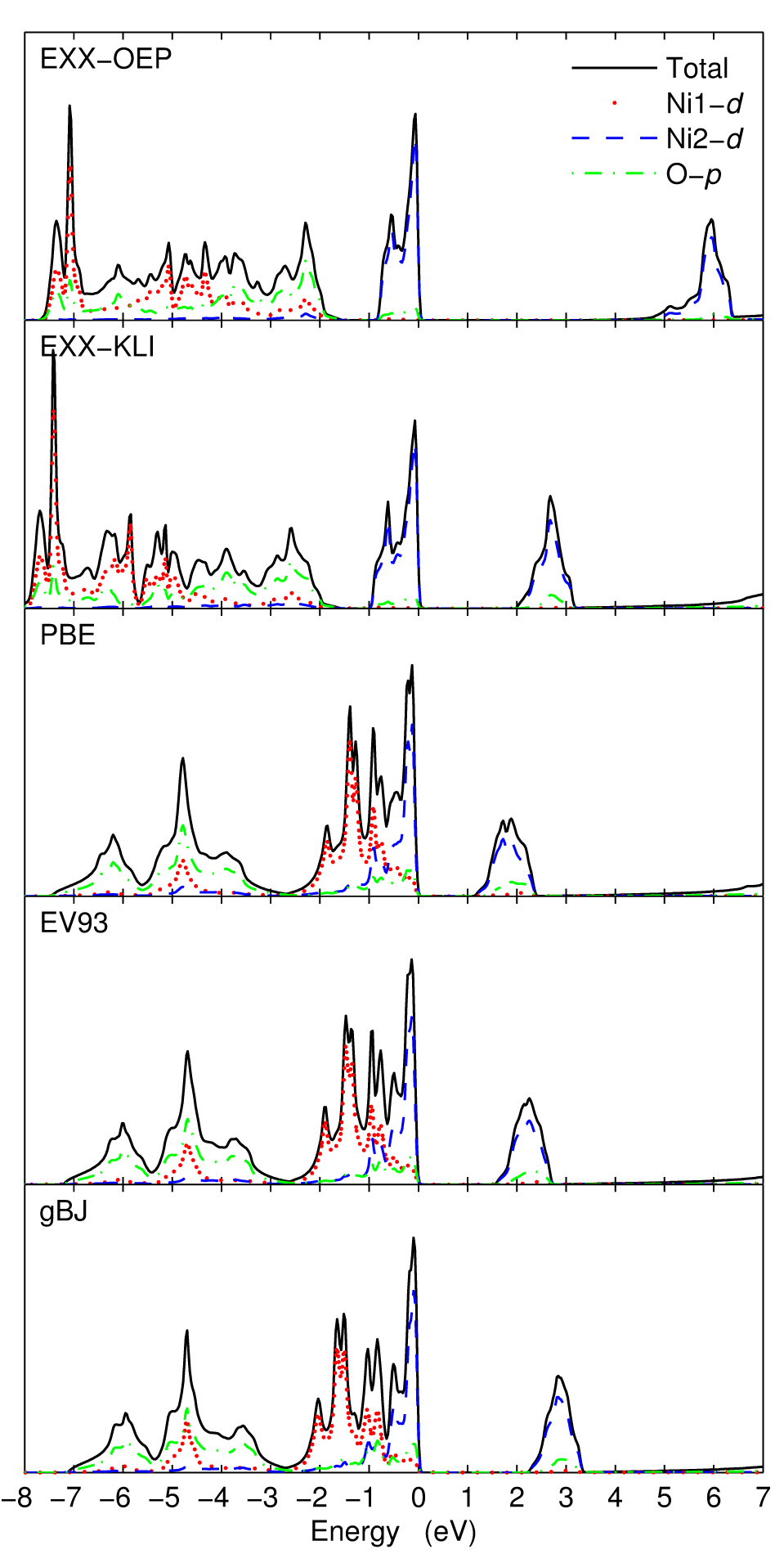}
\caption{\label{fig_NiO_DOS}Spin-up DOS of NiO.
Ni1 is the Ni atom with majority spin-up electrons.
The Fermi energy is set at zero.}
\end{figure}

Besides the KS fundamental band gap, it may also be interesting to look at the
density of states (DOS), in particular for the transition-metal oxides since
qualitative differences in the occupied DOS can be observed. For the other
solids, the visible difference in the DOS consists only of a change in the band
gap, i.e., a rigid shift of the unoccupied states with respect to the occupied
ones. The DOS of antiferromagnetic MnO and NiO are shown in
Figs.~\ref{fig_MnO_DOS} and \ref{fig_NiO_DOS}, respectively.
In MnO, the configuration of the $3d$-electrons on the Mn atom with majority
spin-up electrons is $(t_{2g}^{\uparrow})^{3}(e_{g}^{\uparrow})^{2}
(t_{2g}^{\downarrow})^{0}(e_{g}^{\downarrow})^{0}$ such that the band gap is
determined mainly by the exchange splitting. The EXX-OEP DOS seems overall to be
reproduced more accurately by the EXX-KLI potential. This is clearly the case
for the DOS just below the Fermi energy and the unoccupied DOS, and actually,
the EXX-OEP and EXX-KLI methods describe MnO as an insulator with a band gap of
mixed Mott-Hubbard/charge-transfer type, while the band gap obtained by the
other methods is much more of Mott-Hubbard type. However, in the energy range
between 1 and 7~eV below the Fermi energy, noticeable differences between
EXX-OEP and EXX-KLI can be observed, like for instance the Mn-$3d$ states at
$-2$~eV in the EXX-OEP DOS that are shifted 1 or 2~eV deeper in energy by EXX-KLI.

In NiO, the electronic configuration is $(t_{2g}^{\uparrow})^{3}
(e_{g}^{\uparrow})^{2}(t_{2g}^{\downarrow})^{3}(e_{g}^{\downarrow})^{0}$,
which means a band gap that is determined mainly by the splitting
between the $t_{2g}$ and $e_{g}$ states of the minority spin.
Figure \ref{fig_NiO_DOS} shows that the agreement between EXX-OEP and EXX-KLI
for the DOS is excellent, except for the position of the unoccupied states.
As already observed in Ref.~\onlinecite{TranPRB15}, all semilocal
potentials (including the parameterization of gBJ specific for NiO,
see Fig.~5 of Ref.~\onlinecite{TranPRB15})
lead to DOS which differ significantly from the EXX-OEP DOS,
like showing no sharp Ni-$3d$ peak at the lower part of the valence band or
no clear energy separation between the spin-up and spin-down occupied Ni-$3d$ states.
This is not the case with EXX-KLI, which reproduces accurately
all features in the occupied EXX-OEP DOS.
For the other transition-metal oxides Cu$_{2}$O and ZnO,
the conclusion that the EXX-KLI DOS is the closest to the EXX-OEP remains
also valid.

\begin{figure}[t]
\includegraphics[width=\columnwidth]{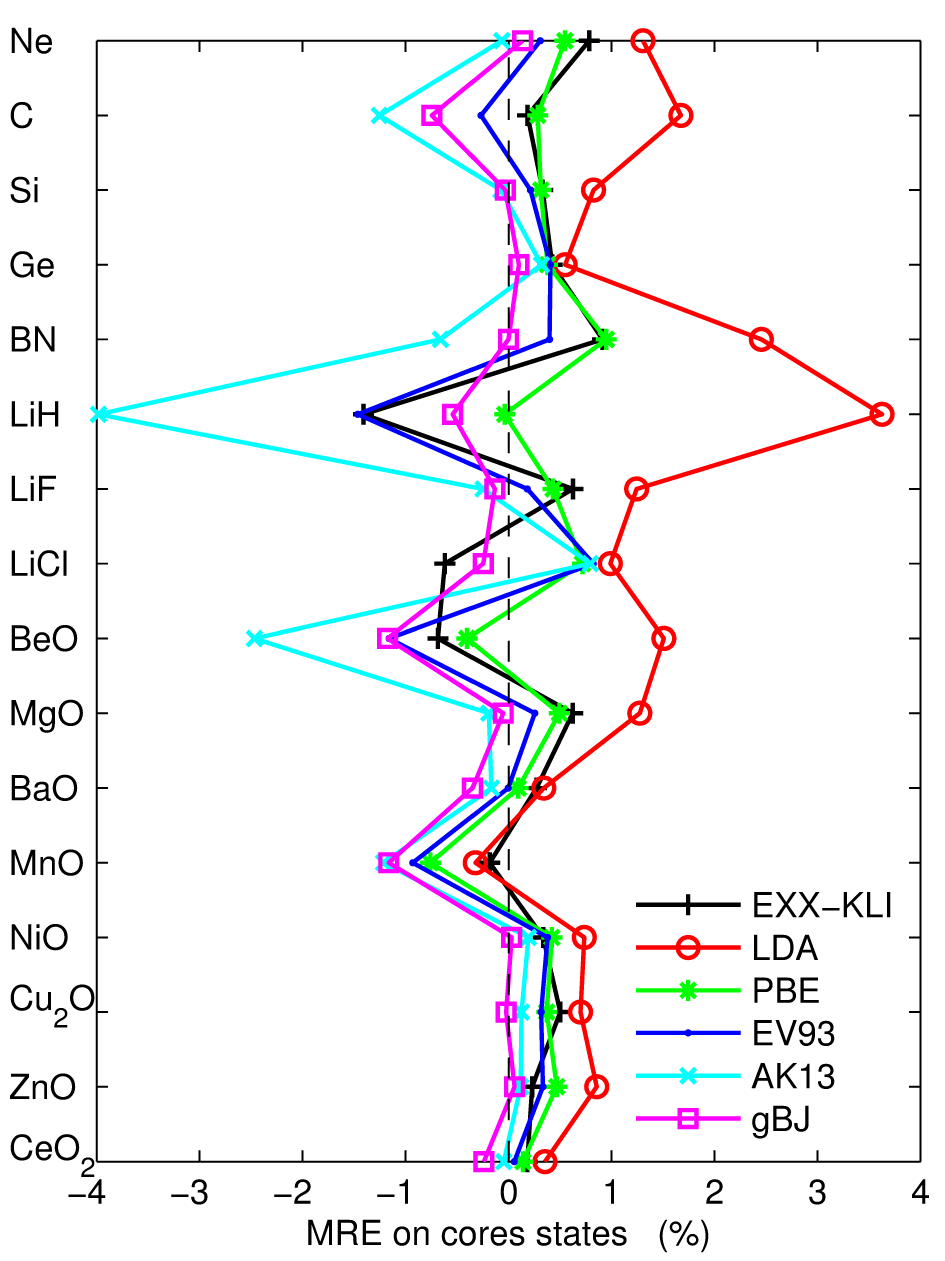}
\caption{\label{fig_core_1}MRE (with respect to EXX-OEP and in \%) for the
energy position of the core states with respect to the VBM.
For a given solid, the MRE is over all core states indicated in Table~S1
(for LiH, the Li-$1s$ state was considered for the present analysis).}
\end{figure}

\begin{figure}[t]
\includegraphics[width=\columnwidth]{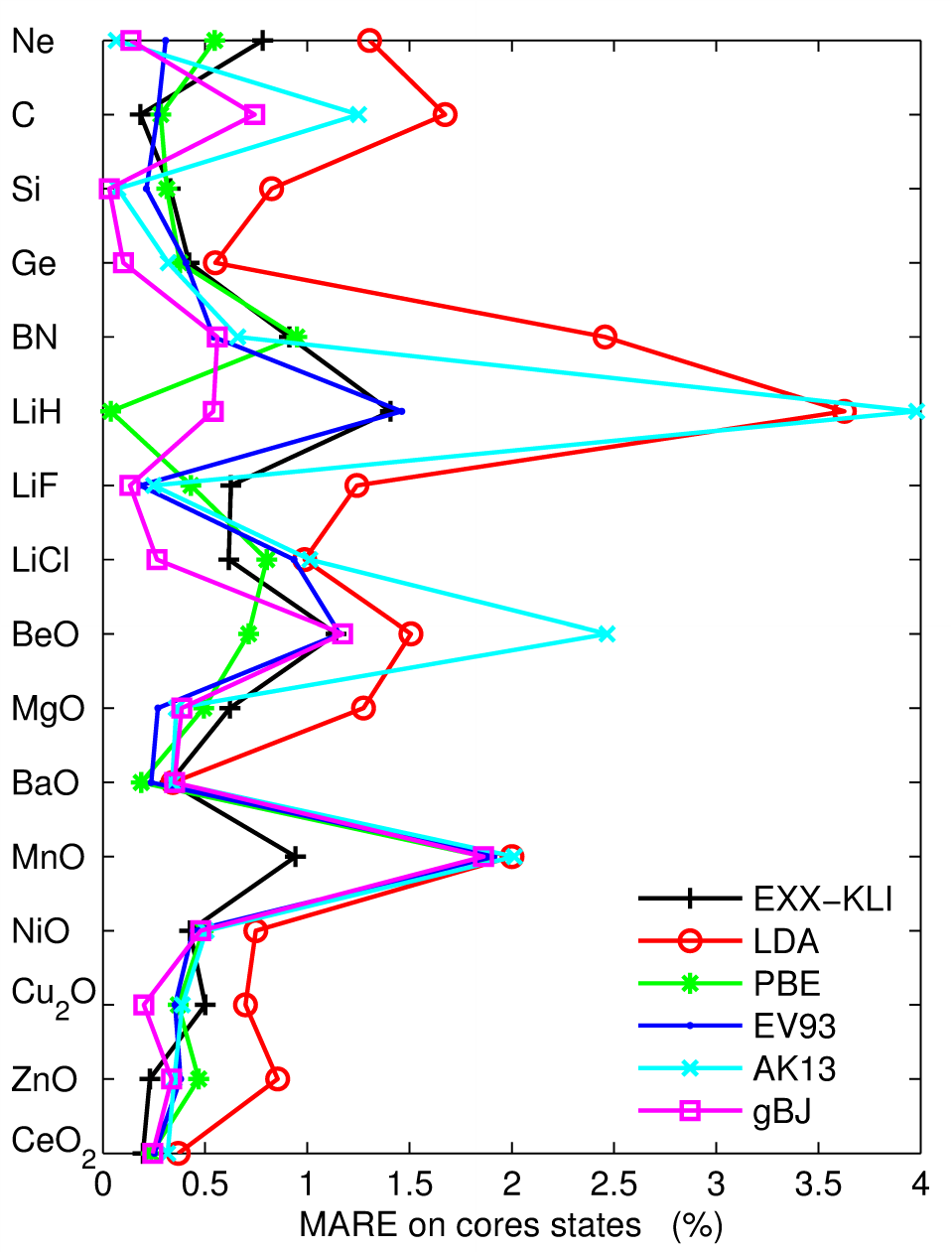}
\caption{\label{fig_core_2}MARE (with respect to EXX-OEP and in \%) for the
energy position of the core states with respect to the VBM.
For a given solid, the MARE is over all core states indicated in Table~S1
(for LiH, the Li-$1s$ state was considered for the present analysis).}
\end{figure}

The results for the energy position of the core states with respect to the
valence band maximum (VBM) are shown in Figs.~\ref{fig_core_1} and \ref{fig_core_2}.
For a given solid and approximate potential, the MRE and MARE (in \%) are defined as
\begin{equation}
\frac{100}{N_{\text{core}}}
\sum_{i=1}^{N_{\text{core}}}\left(
\Delta\varepsilon_{\text{core},i}^{\text{approx}}-
\Delta\varepsilon_{\text{core},i}^{\text{EXX-OEP}}\right)/
\left\vert\Delta\varepsilon_{\text{core},i}^{\text{EXX-OEP}}\right\vert
\label{MREcore}
\end{equation}
and
\begin{equation}
\frac{100}{N_{\text{core}}}
\sum_{i=1}^{N_{\text{core}}}\left\vert
\Delta\varepsilon_{\text{core},i}^{\text{approx}}-
\Delta\varepsilon_{\text{core},i}^{\text{EXX-OEP}}\right\vert
/\left\vert\Delta\varepsilon_{\text{core},i}^{\text{EXX-OEP}}\right\vert,
\label{MAREcore}
\end{equation}
respectively, where the sum runs over the $N_{\text{core}}$ core states
(see Table~S1) and $\Delta\varepsilon_{\text{core},i}$ is the position of
the $i$th core state with respect to the VBM. A negative MRE indicates that on
average the core states are deeper in energy with the approximate potential
than with EXX-OEP. The main observations are the following. On average, LDA
and AK13 lead to too shallow and too deep core states, respectively, since their
mean MRE (MMRE, see Table~\ref{table_statistics}) are 1.1\% and $-0.6\%$. The other
exchange potentials are more accurate and lead to rather similar values
with a MMRE below 0.3\% in magnitude, and a mean MARE (MMARE)
that is in the range 0.5-0.6\%.

\subsection{\label{MMEFG} Magnetic moment and EFG}

\begin{table}
\caption{\label{table_MM_EFG}Atomic spin magnetic moment $\mu_{S}$ (in $\mu_{\text{B}}$)
in MnO and NiO and EFG (in $10^{21}$ V/m$^{2}$) at the Cu site in Cu$_{2}$O
calculated from different exchange potentials.}
\begin{ruledtabular}
\begin{tabular}{lccc}
Potential & $\mu_{S}^{\text{Mn}}$ & $\mu_{S}^{\text{Ni}}$ & EFG$_{\text{Cu}}$ \\
\hline
EXX-OEP & 4.59 & 1.91 & -17.7 \\
EXX-KLI & 4.58 & 1.79 & -11.1 \\
LDA     & 4.18 & 1.30 &  -4.7 \\
PBE     & 4.23 & 1.43 &  -5.6 \\
EV93    & 4.30 & 1.51 &  -6.8 \\
AK13    & 4.39 & 1.58 &  -8.1 \\
gBJ     & 4.35 & 1.61 &  -7.0 \\
HF      & 4.57 & 1.88 & -17.0 \\
\end{tabular}
\end{ruledtabular}
\end{table}

We continue the discussion of the results with the atomic spin magnetic
moment $\mu_{S}$ in MnO and NiO and the EFG in
Cu$_{2}$O. The results in Table~\ref{table_MM_EFG} show that EXX-KLI is a very
good approximation to EXX-OEP for $\mu_{S}$ since the values obtained with
the two methods differ by only $\sim0.1$~$\mu_{\text{B}}$ for NiO and are the same
for MnO. The other exchange potentials lead to substantially smaller values.
We note that in our previous work,\cite{TranPRB15} a value of
1.86~$\mu_{\text{B}}$ for NiO could be obtained with gBJ, but with
parameters $(\gamma,c,p)$ that were tuned specifically for NiO.
The EFG at the Cu site in Cu$_{2}$O has a value of $-17.7\times10^{21}$ V/m$^{2}$ with
EXX-OEP, but is substantially smaller with all other potentials
including EXX-KLI which leads to the best agreement
with $-11.1\times10^{21}$ V/m$^{2}$ ($\sim40\%$ too small).
As for NiO, we could find a parameterization of a modified form of the gBJ
potential (see Ref.~\onlinecite{TranPRB15} for details) that leads to
an EFG approaching the EXX-OEP value.

In addition to the results obtained with the multiplicative exchange potentials, the HF
values are also reported in Table~\ref{table_MM_EFG}, and as already noticed in
Ref.~\onlinecite{TranPRB15}, the EXX-OEP and HF methods provide basically the
same values. This is expected for such properties calculated from the electron
density, since the two methods should in principle lead to electron densities
that should not differ up to the first order,
\cite{KriegerPRA92b,Grabo00,KuemmelPRB03} despite completely different
electronic structures.\cite{TranPRB15}

\subsection{\label{discussion}Further discussion}

\begin{figure}[t]
\includegraphics[width=\columnwidth]{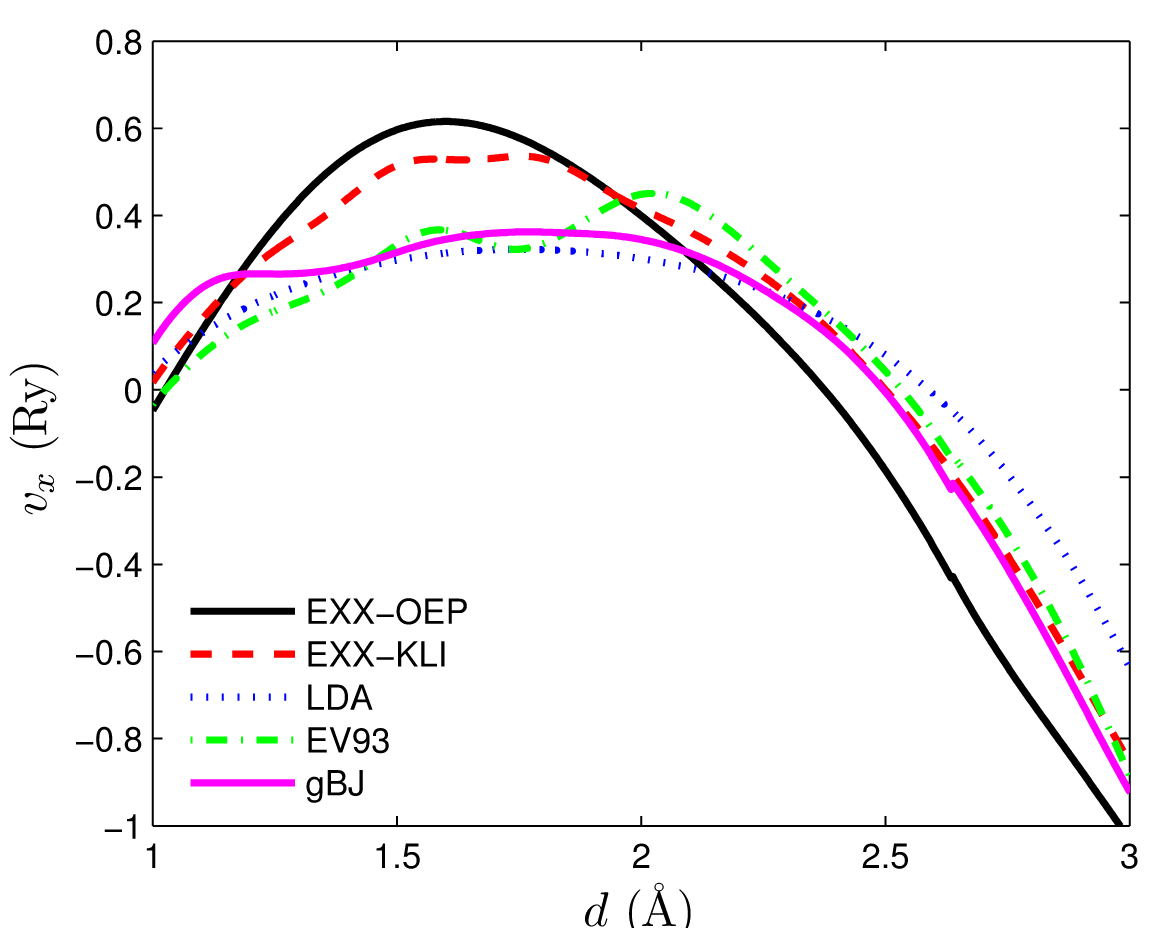}
\caption{\label{fig_Cu2O}
Exchange potentials $v_{\text{x}}$ in Cu$_{2}$O plotted starting at a distance
of 1 \AA~from the Cu atom at site $\left(1/2,1/2,0\right)$ ($d=0$) in the
direction of the O atom at site $\left(3/4,3/4,3/4\right)$ ($d=3.54$~\AA).}
\end{figure}

In our previous works about exchange potentials in solids
\cite{TranJPCM07,BetzingerPRB11,KollerPRB11,TranPRB15,TranJCTC15}
as well as in Refs.~\onlinecite{StaedelePRL97,StaedelePRB99,AulburPRB00,QteishPRB06},
a rather clear understanding of the results could be achieved by visualizing the
potential and electron density. For instance, in solids where the VBM and
conduction band minimum (CBM) are located in different regions of space
(typically, the VBM is localized around atoms and the CBM in the interstitial
region), the size of the band gap is directly related to the
value of the potential in the two regions. The more the values of the
potential in the two regions differ, the more the
band gap should be large (see Ref.~\onlinecite{TranJPCM07} for LiCl and
Ref.~\onlinecite{TranJCTC15} for Kr and BaO).
The situation may be different in transition-metal oxides where
the band gap can be of on-site $d$-$d$ type such that, for instance, it is determined by
the splitting between occupied and unoccupied $d$-states.
In such cases like Cu$_{2}$O\cite{KollerPRB11} or
NiO,\cite{TranPRB15,TranJCTC15} the size of the band gap and
atomic magnetic moment are determined by
the sensitivity of the potential to the $d$-orbital shape (e.g., $t_{2g}$ versus $e_{g}$)
and/or the magnitude of $v_{\text{x},\uparrow}-v_{\text{x},\downarrow}$.
In Ref.~\onlinecite{TranPRB15} it was also shown that the differences between
the electron densities generated by the various potentials correlate quite well
with the numerical results for the total energy, magnetic moment, etc.

From these analyses it was concluded that the LDA and standard GGA potentials
like PBE are much more homogeneous than the EXX-OEP,\cite{TranPRB15}
explaining why they lead
to band gap and magnetic moment that are much smaller than with EXX-OEP.
The more specialized potentials EV93, AK13, and gBJ are more inhomogeneous such
that they are better approximations to the EXX-OEP. This is particularly the
case for the gBJ potential which was shown to reproduce quite accurately most
features of the EXX-OEP, provided that the appropriate parameters
$\gamma$, $c$, and $p$ are used.
The same analysis can also be made for the results obtained in the present
work. However, since the observations and conclusions would be very similar to
those obtained in our previous works, only a brief discussion for two of the
most interesting systems, Cu$_{2}$O and NiO, is given.

Figure~\ref{fig_Cu2O} shows exchange potentials in Cu$_{2}$O plotted along a
portion of the path between the Cu and O atoms located at sites
$\left(1/2,1/2,0\right)$ and $\left(3/4,3/4,3/4\right)$ of the unit cell,
respectively. In Refs.~\onlinecite{TranPRB15,TranJCTC15},
we identified a (valence) region close to the Cu atom
($1\lesssim d\lesssim2$~\AA) to be important for
the band gap and EFG, since it was observed that the potentials
which agree with the EXX-OEP in this region in particular,
namely, gBJ with the universal correction, Becke-Roussel,\cite{BeckePRA88} and
Slater, lead to reasonable values for the band gap and EFG.
To some extent the same is true for the EXX-KLI potential, since from
Fig.~\ref{fig_Cu2O} we can see that it is relatively close to EXX-OEP compared
to the other potentials [see Fig.~8(b) of Ref.~\onlinecite{TranPRB15} and
Fig.~3 of Ref.~\onlinecite{TranJCTC15} for more potentials] and also
leads to smaller difference with respect to EXX-OEP for the
band gap and EFG as discussed above.

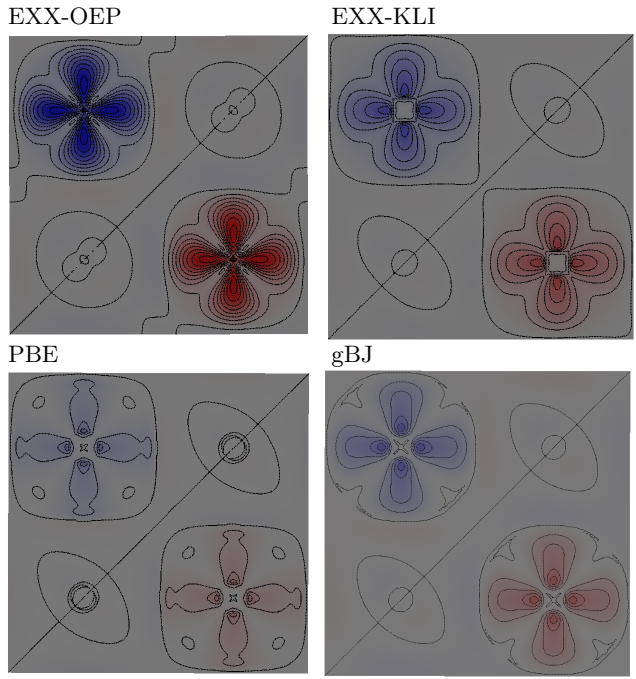
\begin{figure}
\begin{picture}(8.6,9.4)(0,0)
\put(0,4.5){\epsfxsize=4.3cm\epsfbox{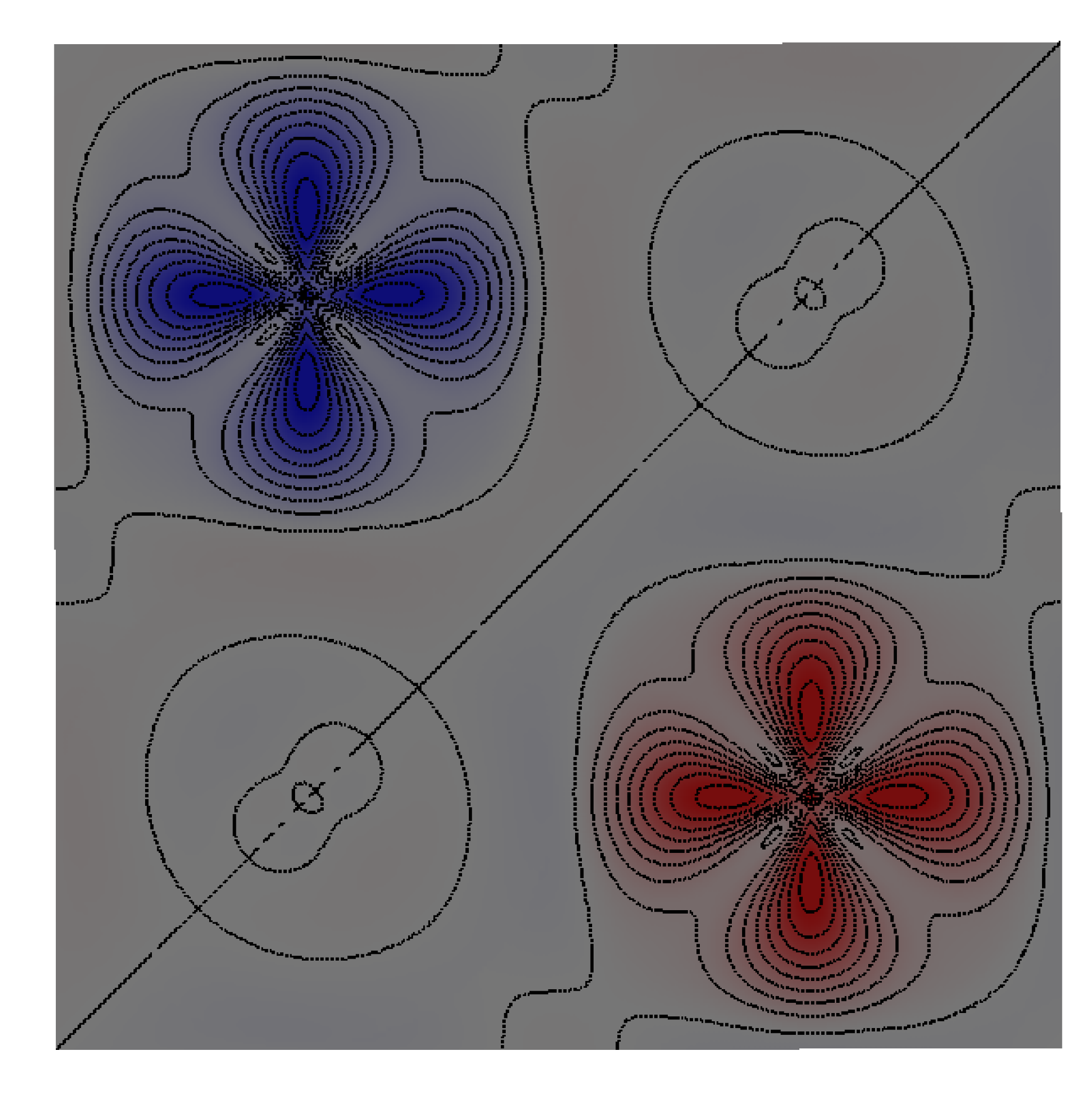}}
\put(4.3,4.5){\epsfxsize=4.3cm\epsfbox{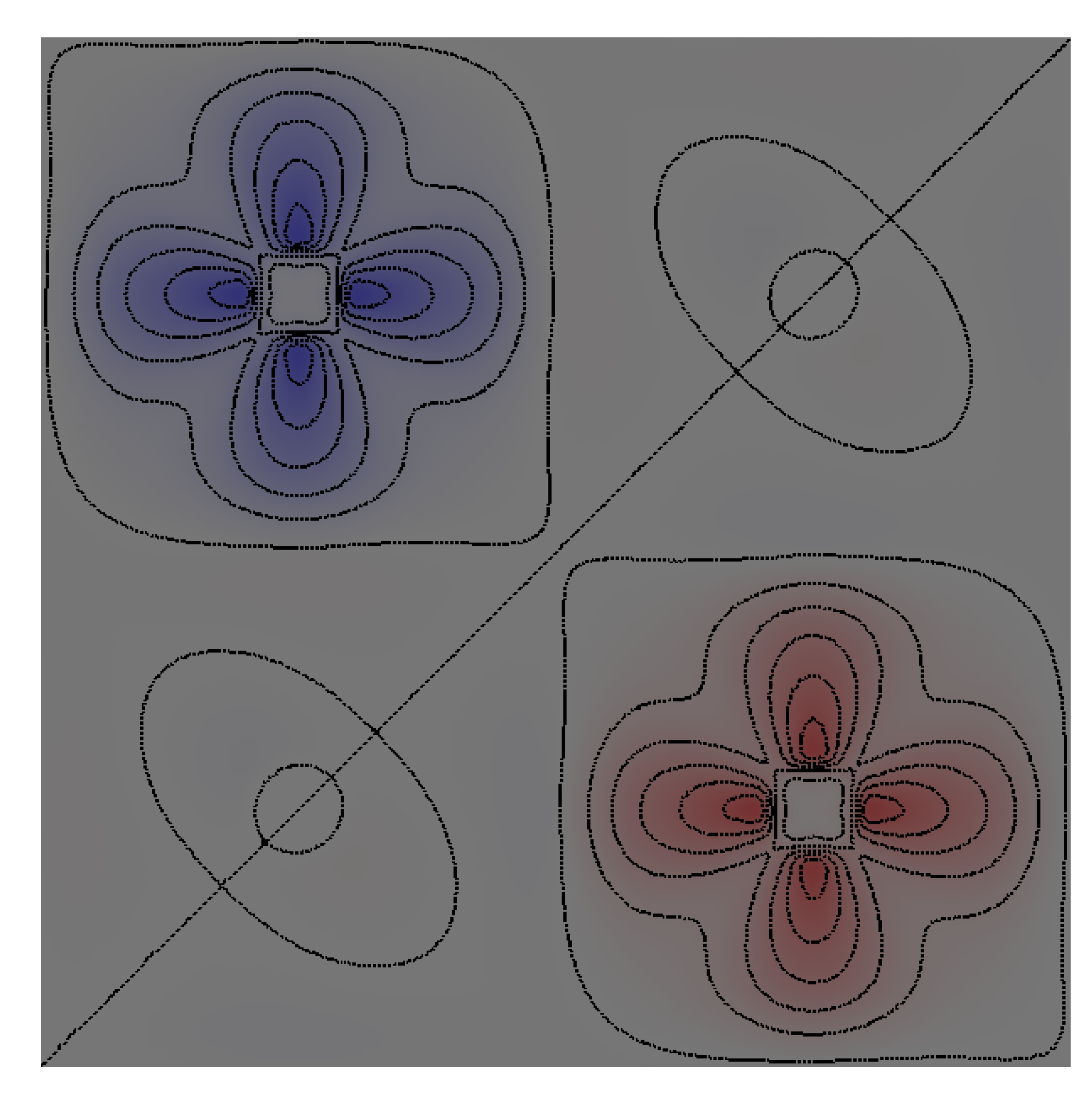}}
\put(0,0){\epsfxsize=4.3cm\epsfbox{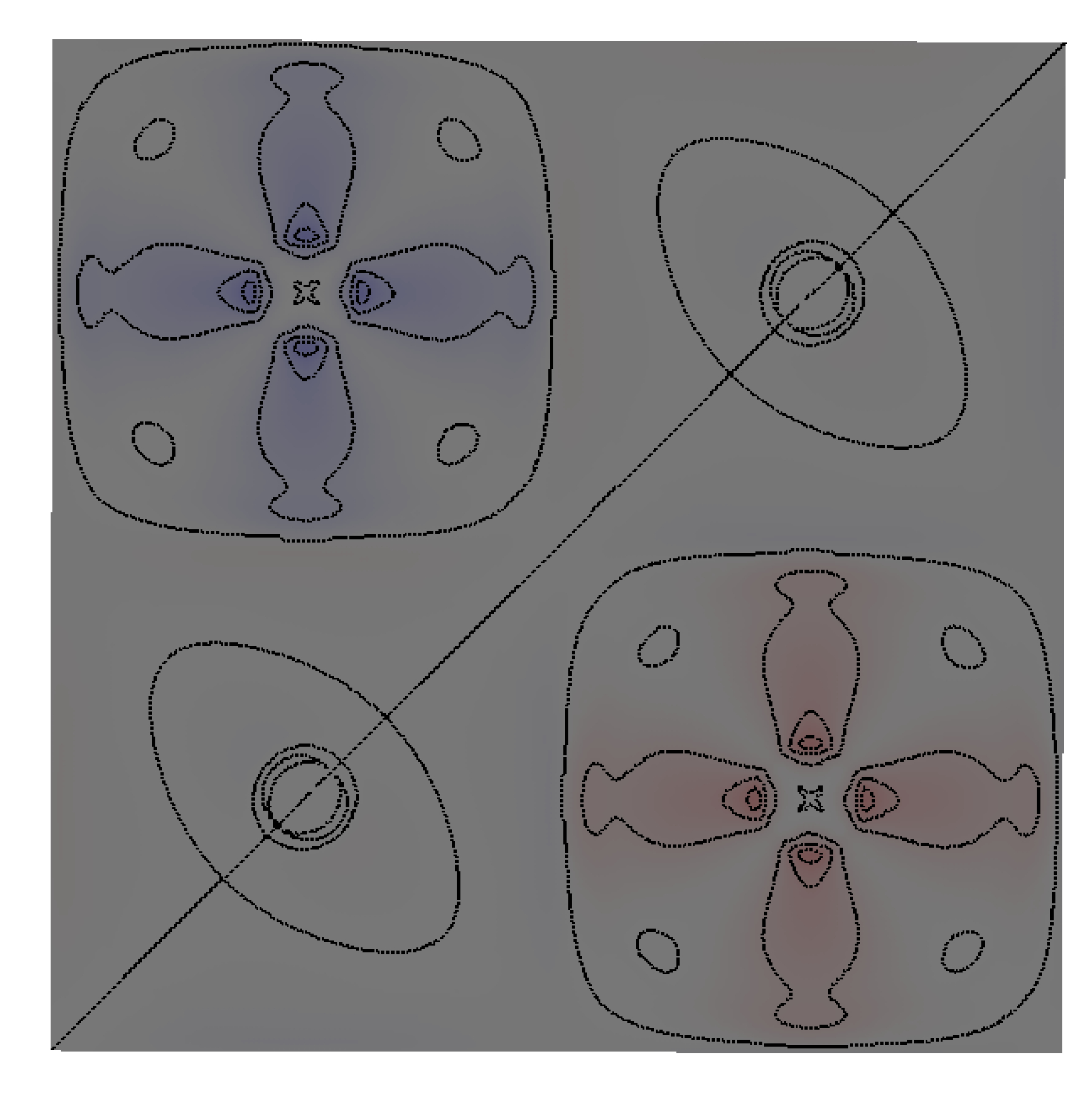}}
\put(4.3,0){\epsfxsize=4.3cm\epsfbox{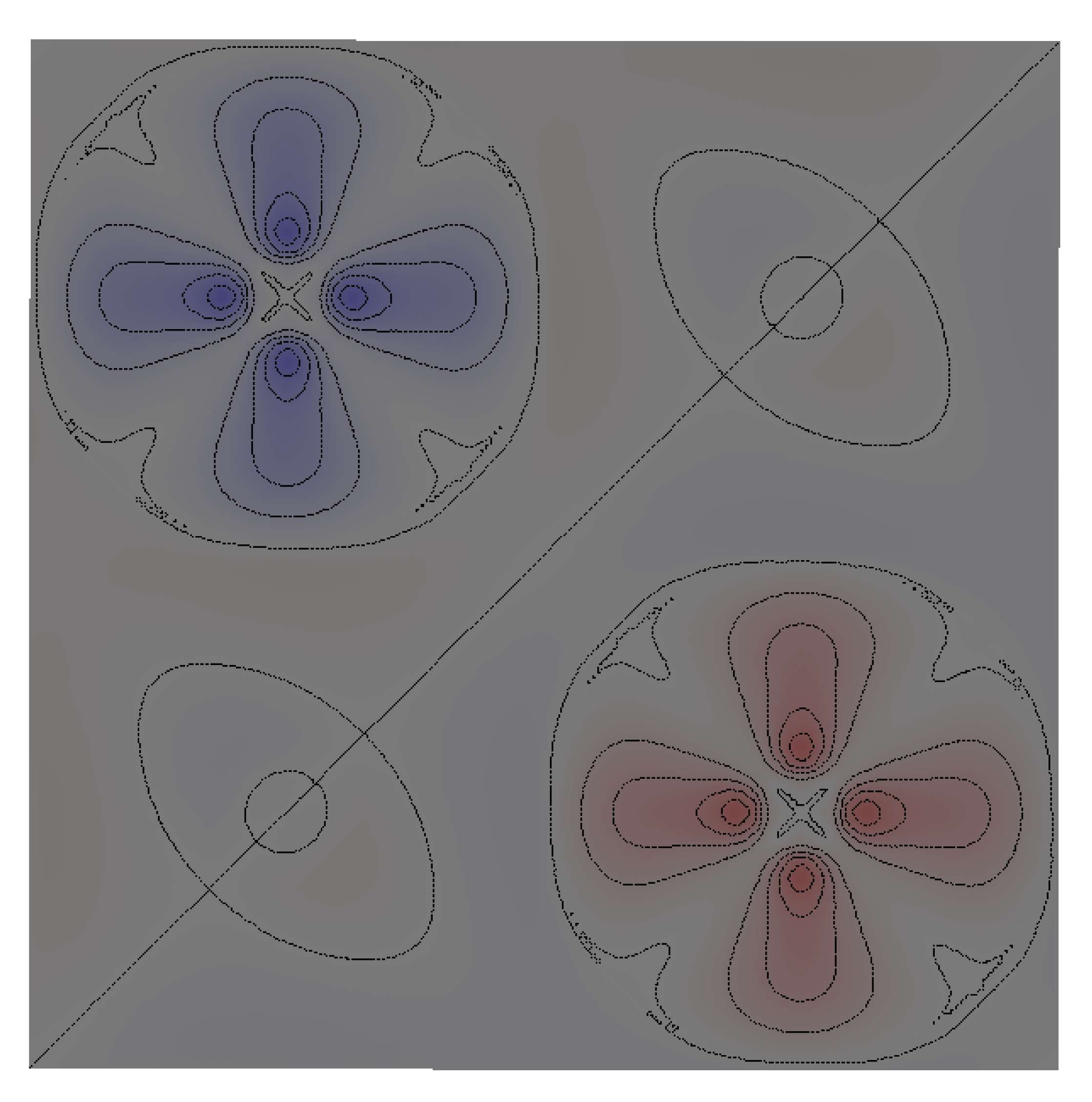}}
\put(0.2,8.8){EXX-OEP}
\put(4.5,8.8){EXX-KLI}
\put(0.2,4.3){PBE}
\put(4.5,4.3){gBJ}
\end{picture}
\caption{\label{fig_NiO}Two-dimensional plots of
$v_{\text{x}}^{\uparrow}-v_{\text{x}}^{\downarrow}$ in a $(001)$ plane of
antiferromagnetic NiO. The contour lines start at $-2$ Ry (blue color) and end
at 2 Ry (red color) with an interval of 0.235 Ry. The Ni atom with a full
spin-up $3d$-shell is at the left upper corner.}
\end{figure}

The difference $v_{\text{x}}^{\uparrow}-v_{\text{x}}^{\downarrow}$
between the spin-up and spin-down exchange potentials for antiferromagnetic NiO
in a (001) plane is shown in Fig.~\ref{fig_NiO}. As we can observe (see Fig.~10
of Ref.~\onlinecite{TranPRB15} and Fig.~4 of Ref.~\onlinecite{TranJCTC15} for
other potentials) the shape of the unoccupied $e_{g}$ orbitals is the most
pronounced with EXX-OEP and all semilocal potentials (except gBJ with
parameters for NiO\cite{TranPRB15}) lead to a $e_{g}$-shape that is very much
attenuated with respect to EXX-OEP. Compared to the semilocal potentials,
EXX-KLI seems to be more accurate, however the magnitude of
$v_{\text{x}}^{\uparrow}-v_{\text{x}}^{\downarrow}$ is still too small, thus
explaining the underestimation of the magnetic moment and band gap.

More generally, since the EXX-KLI potential is derived from the EXX-OEP by using
the closure approximation (i.e., directional averaging), it is expected to be
smoother than the EXX-OEP. This has been underlined by Engel and co-workers in
Refs.~\onlinecite{EngelPRL09,Engel} who already showed that for Si and FeO the
EXX-KLI potential around the atoms is less aspherical than the EXX-OEP.
Thus, for systems with a highly aspherical electron density, e.g., systems with
an open $3d$-shell, the closure approximation should have a large impact on the
results. This is what has indeed been observed for FeO (metallic with EXX-KLI
but not with EXX-OEP\cite{EngelPRL09}) and NiO (much larger underestimation of
the band gap than for the other solids, see Fig.~\ref{fig_gap}).
In comparison, the electron density on the Mn atom in MnO is more spherical
(the $3d$-shell is full for one spin and empty for the other spin), therefore
the underestimation of the band gap is not as large, but similar as for the
non-magnetic solids.

\section{\label{summary}Summary and conclusion}

In this work, we have presented the results of electronic structure
calculations on solids with the EXX-KLI approximation to the exact exchange
potential EXX-OEP. The goals were to provide all-electron benchmark EXX-KLI
(and new EXX-OEP) results and to figure out if EXX-KLI can be used safely as a
substitute to EXX-OEP, and if it is more accurate than the semilocal
approximations like the MGGA gBJ potential. The test set consisted of 16 solids
of various types and the calculated properties were the EXX total energy,
electron density, electronic structure, magnetic moment, and EFG.

The results for the total energy and electronic structure
have shown that \textit{on average} the EXX-KLI and gBJ approximations are
more or less of the same accuracy. However, by looking at the results in
more detail we have noticed that for the transition-metal oxides,
the EXX-KLI and gBJ results can differ qualitatively. For instance, opposite
trends were observed for the band gap in the antiferromagnetic systems;
while EXX-KLI leads to a fairly accurate band gap in MnO (clearly more accurate
than gBJ), it is by far too small or even zero for NiO, CoO, and FeO
(gBJ is better than EXX-KLI for these cases). The EXX-KLI approximation
seems to be quite inaccurate in the case of highly aspherical electron density
like in NiO, FeO, and CoO as noticed previously.\cite{EngelPRL09}
On the other hand, the EXX-OEP occupied DOS of MnO and
NiO are reproduced accurately by EXX-KLI, while all semilocal potentials lead to
completely different DOS, especially for NiO.
The other difference between EXX-KLI and gBJ is the error for the band gap:
with EXX-KLI there is a systematic underestimation of the order of $\sim0.5$~eV for
all systems except NiO, while for gBJ and all other semilocal potentials the error
varies strongly among the compounds.

For the magnetic moment and EFG, the EXX-OEP results are reproduced
more accurately by EXX-KLI, nevertheless a clear underestimation of the magnitude
of the EFG in Cu$_{2}$O is still observed.

Thus, in conclusion, EXX-KLI seems to be a rather good approximation to EXX-OEP
for ground-state properties, i.e., properties which are calculated using the
occupied orbitals. For the band gap, an excited-state property, EXX-KLI leads
to an underestimation of $\sim0.5$~eV for most systems, except in
the special case of antiferromagnetic NiO (and also FeO and CoO) for which a much
larger error of more than 1.5~eV is obtained. The results obtained with gBJ,
the most accurate of the tested semilocal potentials, are also rather good,
but more unpredictable for the band gap, a behavior which is in general
more expected for semilocal approximations than for \textit{ab initio}
approximations like EXX-KLI.

Concerning the LHF/CEDA\cite{DellaSalaJCP01,GritsenkoPRA01}
method briefly mentioned in Sec.~\ref{theory}, which, in principle, should be
a better approximation to EXX-OEP (but also more expensive) than KLI, the works
published so far\cite{DellaSalaJCP01,GruningJCP02}
have shown that the LHF/CEDA and KLI results for the total energy and gap
are quasi-identical in most cases (see also Ref. \onlinecite{KuemmelRMP08} for
further discussion). However, since these LHF/CEDA calculations were
done for atoms and light molecules, it is not certain that this conclusion
would hold also for much more complicated systems like NiO or FeO.

\begin{acknowledgments}

This work was supported by the project SFB-F41 (ViCoM) of the Austrian Science
Fund. M. B. gratefully acknowledges financial support from the Helmholtz
Association through the Hemholtz Postdoc Programme (VH-PD-022).
We are grateful to Eberhard Engel for very useful discussions.

\end{acknowledgments}

\bibliography{/planck/tran/divers/references}

\begin{thebibliography}{93}%
\makeatletter
\providecommand \@ifxundefined [1]{%
 \@ifx{#1\undefined}
}%
\providecommand \@ifnum [1]{%
 \ifnum #1\expandafter \@firstoftwo
 \else \expandafter \@secondoftwo
 \fi
}%
\providecommand \@ifx [1]{%
 \ifx #1\expandafter \@firstoftwo
 \else \expandafter \@secondoftwo
 \fi
}%
\providecommand \natexlab [1]{#1}%
\providecommand \enquote  [1]{``#1''}%
\providecommand \bibnamefont  [1]{#1}%
\providecommand \bibfnamefont [1]{#1}%
\providecommand \citenamefont [1]{#1}%
\providecommand \href@noop [0]{\@secondoftwo}%
\providecommand \href [0]{\begingroup \@sanitize@url \@href}%
\providecommand \@href[1]{\@@startlink{#1}\@@href}%
\providecommand \@@href[1]{\endgroup#1\@@endlink}%
\providecommand \@sanitize@url [0]{\catcode `\\12\catcode `\$12\catcode
  `\&12\catcode `\#12\catcode `\^12\catcode `\_12\catcode `\%12\relax}%
\providecommand \@@startlink[1]{}%
\providecommand \@@endlink[0]{}%
\providecommand \url  [0]{\begingroup\@sanitize@url \@url }%
\providecommand \@url [1]{\endgroup\@href {#1}{\urlprefix }}%
\providecommand \urlprefix  [0]{URL }%
\providecommand \Eprint [0]{\href }%
\providecommand \doibase [0]{http://dx.doi.org/}%
\providecommand \selectlanguage [0]{\@gobble}%
\providecommand \bibinfo  [0]{\@secondoftwo}%
\providecommand \bibfield  [0]{\@secondoftwo}%
\providecommand \translation [1]{[#1]}%
\providecommand \BibitemOpen [0]{}%
\providecommand \bibitemStop [0]{}%
\providecommand \bibitemNoStop [0]{.\EOS\space}%
\providecommand \EOS [0]{\spacefactor3000\relax}%
\providecommand \BibitemShut  [1]{\csname bibitem#1\endcsname}%
\let\auto@bib@innerbib\@empty
\bibitem [{\citenamefont {Hohenberg}\ and\ \citenamefont
  {Kohn}(1964)}]{HohenbergPR64}%
  \BibitemOpen
  \bibfield  {author} {\bibinfo {author} {\bibfnamefont {P.}~\bibnamefont
  {Hohenberg}}\ and\ \bibinfo {author} {\bibfnamefont {W.}~\bibnamefont
  {Kohn}},\ }\href@noop {} {\bibfield  {journal} {\bibinfo  {journal} {Phys.
  Rev.}\ }\textbf {\bibinfo {volume} {136}},\ \bibinfo {pages} {B864} (\bibinfo
  {year} {1964})}\BibitemShut {NoStop}%
\bibitem [{\citenamefont {Kohn}\ and\ \citenamefont {Sham}(1965)}]{KohnPR65}%
  \BibitemOpen
  \bibfield  {author} {\bibinfo {author} {\bibfnamefont {W.}~\bibnamefont
  {Kohn}}\ and\ \bibinfo {author} {\bibfnamefont {L.~J.}\ \bibnamefont
  {Sham}},\ }\href@noop {} {\bibfield  {journal} {\bibinfo  {journal} {Phys.
  Rev.}\ }\textbf {\bibinfo {volume} {140}},\ \bibinfo {pages} {A1133}
  (\bibinfo {year} {1965})}\BibitemShut {NoStop}%
\bibitem [{\citenamefont {Kim}\ \emph {et~al.}(2013)\citenamefont {Kim},
  \citenamefont {Sim},\ and\ \citenamefont {Burke}}]{KimPRL13}%
  \BibitemOpen
  \bibfield  {author} {\bibinfo {author} {\bibfnamefont {M.-C.}\ \bibnamefont
  {Kim}}, \bibinfo {author} {\bibfnamefont {E.}~\bibnamefont {Sim}}, \ and\
  \bibinfo {author} {\bibfnamefont {K.}~\bibnamefont {Burke}},\ }\href@noop {}
  {\bibfield  {journal} {\bibinfo  {journal} {Phys. Rev. Lett.}\ }\textbf
  {\bibinfo {volume} {111}},\ \bibinfo {pages} {073003} (\bibinfo {year}
  {2013})}\BibitemShut {NoStop}%
\bibitem [{\citenamefont {K{\"u}mmel}\ and\ \citenamefont
  {Kronik}(2008)}]{KuemmelRMP08}%
  \BibitemOpen
  \bibfield  {author} {\bibinfo {author} {\bibfnamefont {S.}~\bibnamefont
  {K{\"u}mmel}}\ and\ \bibinfo {author} {\bibfnamefont {L.}~\bibnamefont
  {Kronik}},\ }\href@noop {} {\bibfield  {journal} {\bibinfo  {journal} {Rev.
  Mod. Phys.}\ }\textbf {\bibinfo {volume} {80}},\ \bibinfo {pages} {3}
  (\bibinfo {year} {2008})}\BibitemShut {NoStop}%
\bibitem [{\citenamefont {Cohen}\ \emph {et~al.}(2012)\citenamefont {Cohen},
  \citenamefont {Mori-S{\'{a}}nchez},\ and\ \citenamefont {Yang}}]{CohenCR12}%
  \BibitemOpen
  \bibfield  {author} {\bibinfo {author} {\bibfnamefont {A.~J.}\ \bibnamefont
  {Cohen}}, \bibinfo {author} {\bibfnamefont {P.}~\bibnamefont
  {Mori-S{\'{a}}nchez}}, \ and\ \bibinfo {author} {\bibfnamefont
  {W.}~\bibnamefont {Yang}},\ }\href@noop {} {\bibfield  {journal} {\bibinfo
  {journal} {Chem. Rev.}\ }\textbf {\bibinfo {volume} {112}},\ \bibinfo {pages}
  {289} (\bibinfo {year} {2012})}\BibitemShut {NoStop}%
\bibitem [{\citenamefont {Seidl}\ \emph {et~al.}(1996)\citenamefont {Seidl},
  \citenamefont {G\"orling}, \citenamefont {Vogl}, \citenamefont {Majewski},\
  and\ \citenamefont {Levy}}]{SeidlPRB96}%
  \BibitemOpen
  \bibfield  {author} {\bibinfo {author} {\bibfnamefont {A.}~\bibnamefont
  {Seidl}}, \bibinfo {author} {\bibfnamefont {A.}~\bibnamefont {G\"orling}},
  \bibinfo {author} {\bibfnamefont {P.}~\bibnamefont {Vogl}}, \bibinfo {author}
  {\bibfnamefont {J.~A.}\ \bibnamefont {Majewski}}, \ and\ \bibinfo {author}
  {\bibfnamefont {M.}~\bibnamefont {Levy}},\ }\href@noop {} {\bibfield
  {journal} {\bibinfo  {journal} {Phys. Rev. B}\ }\textbf {\bibinfo {volume}
  {53}},\ \bibinfo {pages} {3764} (\bibinfo {year} {1996})}\BibitemShut
  {NoStop}%
\bibitem [{\citenamefont {Sharp}\ and\ \citenamefont
  {Horton}(1953)}]{SharpPR53}%
  \BibitemOpen
  \bibfield  {author} {\bibinfo {author} {\bibfnamefont {R.~T.}\ \bibnamefont
  {Sharp}}\ and\ \bibinfo {author} {\bibfnamefont {G.~K.}\ \bibnamefont
  {Horton}},\ }\href@noop {} {\bibfield  {journal} {\bibinfo  {journal} {Phys.
  Rev.}\ }\textbf {\bibinfo {volume} {90}},\ \bibinfo {pages} {317} (\bibinfo
  {year} {1953})}\BibitemShut {NoStop}%
\bibitem [{\citenamefont {Engel}\ and\ \citenamefont {Dreizler}(2011)}]{Engel}%
  \BibitemOpen
  \bibfield  {author} {\bibinfo {author} {\bibfnamefont {E.}~\bibnamefont
  {Engel}}\ and\ \bibinfo {author} {\bibfnamefont {R.~M.}\ \bibnamefont
  {Dreizler}},\ }\href@noop {} {\emph {\bibinfo {title} {Density Functional
  Theory: An Advanced Course}}}\ (\bibinfo  {publisher} {Springer},\ \bibinfo
  {address} {Berlin},\ \bibinfo {year} {2011})\BibitemShut {NoStop}%
\bibitem [{\citenamefont {Betzinger}\ \emph {et~al.}(2011)\citenamefont
  {Betzinger}, \citenamefont {Friedrich}, \citenamefont {Bl\"ugel},\ and\
  \citenamefont {G\"orling}}]{BetzingerPRB11}%
  \BibitemOpen
  \bibfield  {author} {\bibinfo {author} {\bibfnamefont {M.}~\bibnamefont
  {Betzinger}}, \bibinfo {author} {\bibfnamefont {C.}~\bibnamefont
  {Friedrich}}, \bibinfo {author} {\bibfnamefont {S.}~\bibnamefont {Bl\"ugel}},
  \ and\ \bibinfo {author} {\bibfnamefont {A.}~\bibnamefont {G\"orling}},\
  }\href@noop {} {\bibfield  {journal} {\bibinfo  {journal} {Phys. Rev. B}\
  }\textbf {\bibinfo {volume} {83}},\ \bibinfo {pages} {045105} (\bibinfo
  {year} {2011})}\BibitemShut {NoStop}%
\bibitem [{\citenamefont {Gidopoulos}\ and\ \citenamefont
  {Lathiotakis}(2012)}]{GidopoulosPRA12}%
  \BibitemOpen
  \bibfield  {author} {\bibinfo {author} {\bibfnamefont {N.~I.}\ \bibnamefont
  {Gidopoulos}}\ and\ \bibinfo {author} {\bibfnamefont {N.~N.}\ \bibnamefont
  {Lathiotakis}},\ }\href@noop {} {\bibfield  {journal} {\bibinfo  {journal}
  {Phys. Rev. A}\ }\textbf {\bibinfo {volume} {85}},\ \bibinfo {pages} {052508}
  (\bibinfo {year} {2012})};\ \bibinfo {note} {\textbf{88}, 046502
  (2013)}\BibitemShut {NoStop}%
\bibitem [{\citenamefont {Engel}(2014{\natexlab{a}})}]{EngelJCP14}%
  \BibitemOpen
  \bibfield  {author} {\bibinfo {author} {\bibfnamefont {E.}~\bibnamefont
  {Engel}},\ }\href@noop {} {\bibfield  {journal} {\bibinfo  {journal} {J.
  Chem. Phys.}\ }\textbf {\bibinfo {volume} {140}},\ \bibinfo {pages} {18A505}
  (\bibinfo {year} {2014}{\natexlab{a}})}\BibitemShut {NoStop}%
\bibitem [{\citenamefont {Tran}\ \emph
  {et~al.}(2015{\natexlab{a}})\citenamefont {Tran}, \citenamefont {Blaha},
  \citenamefont {Betzinger},\ and\ \citenamefont {Bl\"{u}gel}}]{TranPRB15}%
  \BibitemOpen
  \bibfield  {author} {\bibinfo {author} {\bibfnamefont {F.}~\bibnamefont
  {Tran}}, \bibinfo {author} {\bibfnamefont {P.}~\bibnamefont {Blaha}},
  \bibinfo {author} {\bibfnamefont {M.}~\bibnamefont {Betzinger}}, \ and\
  \bibinfo {author} {\bibfnamefont {S.}~\bibnamefont {Bl\"{u}gel}},\
  }\href@noop {} {\bibfield  {journal} {\bibinfo  {journal} {Phys. Rev. B}\
  }\textbf {\bibinfo {volume} {91}},\ \bibinfo {pages} {165121} (\bibinfo
  {year} {2015}{\natexlab{a}})}\BibitemShut {NoStop}%
\bibitem [{\citenamefont {Becke}\ and\ \citenamefont
  {Johnson}(2006)}]{BeckeJCP06}%
  \BibitemOpen
  \bibfield  {author} {\bibinfo {author} {\bibfnamefont {A.~D.}\ \bibnamefont
  {Becke}}\ and\ \bibinfo {author} {\bibfnamefont {E.~R.}\ \bibnamefont
  {Johnson}},\ }\href@noop {} {\bibfield  {journal} {\bibinfo  {journal} {J.
  Chem. Phys.}\ }\textbf {\bibinfo {volume} {124}},\ \bibinfo {pages} {221101}
  (\bibinfo {year} {2006})}\BibitemShut {NoStop}%
\bibitem [{\citenamefont {Krieger}\ \emph {et~al.}(1990)\citenamefont
  {Krieger}, \citenamefont {Li},\ and\ \citenamefont
  {Iafrate}}]{KriegerPLA90a}%
  \BibitemOpen
  \bibfield  {author} {\bibinfo {author} {\bibfnamefont {J.~B.}\ \bibnamefont
  {Krieger}}, \bibinfo {author} {\bibfnamefont {Y.}~\bibnamefont {Li}}, \ and\
  \bibinfo {author} {\bibfnamefont {G.~J.}\ \bibnamefont {Iafrate}},\
  }\href@noop {} {\bibfield  {journal} {\bibinfo  {journal} {Phys. Lett. A}\
  }\textbf {\bibinfo {volume} {146}},\ \bibinfo {pages} {256} (\bibinfo {year}
  {1990})}\BibitemShut {NoStop}%
\bibitem [{\citenamefont {Krieger}\ \emph
  {et~al.}(1992{\natexlab{a}})\citenamefont {Krieger}, \citenamefont {Li},\
  and\ \citenamefont {Iafrate}}]{KriegerPRA92a}%
  \BibitemOpen
  \bibfield  {author} {\bibinfo {author} {\bibfnamefont {J.~B.}\ \bibnamefont
  {Krieger}}, \bibinfo {author} {\bibfnamefont {Y.}~\bibnamefont {Li}}, \ and\
  \bibinfo {author} {\bibfnamefont {G.~J.}\ \bibnamefont {Iafrate}},\
  }\href@noop {} {\bibfield  {journal} {\bibinfo  {journal} {Phys. Rev. A}\
  }\textbf {\bibinfo {volume} {45}},\ \bibinfo {pages} {101} (\bibinfo {year}
  {1992}{\natexlab{a}})}\BibitemShut {NoStop}%
\bibitem [{\citenamefont {Krieger}\ \emph
  {et~al.}(1992{\natexlab{b}})\citenamefont {Krieger}, \citenamefont {Li},\
  and\ \citenamefont {Iafrate}}]{KriegerPRA92b}%
  \BibitemOpen
  \bibfield  {author} {\bibinfo {author} {\bibfnamefont {J.~B.}\ \bibnamefont
  {Krieger}}, \bibinfo {author} {\bibfnamefont {Y.}~\bibnamefont {Li}}, \ and\
  \bibinfo {author} {\bibfnamefont {G.~J.}\ \bibnamefont {Iafrate}},\
  }\href@noop {} {\bibfield  {journal} {\bibinfo  {journal} {Phys. Rev. A}\
  }\textbf {\bibinfo {volume} {46}},\ \bibinfo {pages} {5453} (\bibinfo {year}
  {1992}{\natexlab{b}})}\BibitemShut {NoStop}%
\bibitem [{\citenamefont {Li}\ \emph {et~al.}(1993)\citenamefont {Li},
  \citenamefont {Krieger},\ and\ \citenamefont {Iafrate}}]{LiPRA93}%
  \BibitemOpen
  \bibfield  {author} {\bibinfo {author} {\bibfnamefont {Y.}~\bibnamefont
  {Li}}, \bibinfo {author} {\bibfnamefont {J.~B.}\ \bibnamefont {Krieger}}, \
  and\ \bibinfo {author} {\bibfnamefont {G.~J.}\ \bibnamefont {Iafrate}},\
  }\href@noop {} {\bibfield  {journal} {\bibinfo  {journal} {Phys. Rev. A}\
  }\textbf {\bibinfo {volume} {47}},\ \bibinfo {pages} {165} (\bibinfo {year}
  {1993})}\BibitemShut {NoStop}%
\bibitem [{\citenamefont {Li}\ \emph {et~al.}(1991)\citenamefont {Li},
  \citenamefont {Krieger}, \citenamefont {Norman},\ and\ \citenamefont
  {Iafrate}}]{LiPRB91}%
  \BibitemOpen
  \bibfield  {author} {\bibinfo {author} {\bibfnamefont {Y.}~\bibnamefont
  {Li}}, \bibinfo {author} {\bibfnamefont {J.~B.}\ \bibnamefont {Krieger}},
  \bibinfo {author} {\bibfnamefont {M.~R.}\ \bibnamefont {Norman}}, \ and\
  \bibinfo {author} {\bibfnamefont {G.~J.}\ \bibnamefont {Iafrate}},\
  }\href@noop {} {\bibfield  {journal} {\bibinfo  {journal} {Phys. Rev. B}\
  }\textbf {\bibinfo {volume} {44}},\ \bibinfo {pages} {10437} (\bibinfo {year}
  {1991})}\BibitemShut {NoStop}%
\bibitem [{\citenamefont {Tong}\ and\ \citenamefont {Chu}(1997)}]{TongPRA97}%
  \BibitemOpen
  \bibfield  {author} {\bibinfo {author} {\bibfnamefont {X.-M.}\ \bibnamefont
  {Tong}}\ and\ \bibinfo {author} {\bibfnamefont {S.-I}\ \bibnamefont {Chu}},\
  }\href@noop {} {\bibfield  {journal} {\bibinfo  {journal} {Phys. Rev. A}\
  }\textbf {\bibinfo {volume} {55}},\ \bibinfo {pages} {3406} (\bibinfo {year}
  {1997})}\BibitemShut {NoStop}%
\bibitem [{\citenamefont {Garza}\ \emph {et~al.}(2000)\citenamefont {Garza},
  \citenamefont {Nichols},\ and\ \citenamefont {Dixon}}]{GarzaJCP00}%
  \BibitemOpen
  \bibfield  {author} {\bibinfo {author} {\bibfnamefont {J.}~\bibnamefont
  {Garza}}, \bibinfo {author} {\bibfnamefont {J.~A.}\ \bibnamefont {Nichols}},
  \ and\ \bibinfo {author} {\bibfnamefont {D.~A.}\ \bibnamefont {Dixon}},\
  }\href@noop {} {\bibfield  {journal} {\bibinfo  {journal} {J. Chem. Phys.}\
  }\textbf {\bibinfo {volume} {112}},\ \bibinfo {pages} {7880} (\bibinfo {year}
  {2000})}\BibitemShut {NoStop}%
\bibitem [{\citenamefont {Patchkovskii}\ \emph {et~al.}(2001)\citenamefont
  {Patchkovskii}, \citenamefont {Autschbach},\ and\ \citenamefont
  {Ziegler}}]{PatchkovskiiJCP01}%
  \BibitemOpen
  \bibfield  {author} {\bibinfo {author} {\bibfnamefont {S.}~\bibnamefont
  {Patchkovskii}}, \bibinfo {author} {\bibfnamefont {J.}~\bibnamefont
  {Autschbach}}, \ and\ \bibinfo {author} {\bibfnamefont {T.}~\bibnamefont
  {Ziegler}},\ }\href@noop {} {\bibfield  {journal} {\bibinfo  {journal} {J.
  Chem. Phys.}\ }\textbf {\bibinfo {volume} {115}},\ \bibinfo {pages} {26}
  (\bibinfo {year} {2001})}\BibitemShut {NoStop}%
\bibitem [{\citenamefont {Arbuznikov}\ and\ \citenamefont
  {Kaupp}(2003)}]{ArbuznikovCPL03}%
  \BibitemOpen
  \bibfield  {author} {\bibinfo {author} {\bibfnamefont {A.~V.}\ \bibnamefont
  {Arbuznikov}}\ and\ \bibinfo {author} {\bibfnamefont {M.}~\bibnamefont
  {Kaupp}},\ }\href@noop {} {\bibfield  {journal} {\bibinfo  {journal} {Chem.
  Phys. Lett.}\ }\textbf {\bibinfo {volume} {381}},\ \bibinfo {pages} {495}
  (\bibinfo {year} {2003})}\BibitemShut {NoStop}%
\bibitem [{\citenamefont {Eich}\ and\ \citenamefont
  {Hellgren}(2014)}]{EichJCP14}%
  \BibitemOpen
  \bibfield  {author} {\bibinfo {author} {\bibfnamefont {F.~G.}\ \bibnamefont
  {Eich}}\ and\ \bibinfo {author} {\bibfnamefont {M.}~\bibnamefont
  {Hellgren}},\ }\href@noop {} {\bibfield  {journal} {\bibinfo  {journal} {J.
  Chem. Phys.}\ }\textbf {\bibinfo {volume} {141}},\ \bibinfo {pages} {224107}
  (\bibinfo {year} {2014})}\BibitemShut {NoStop}%
\bibitem [{\citenamefont {Yang}\ \emph {et~al.}(2016)\citenamefont {Yang},
  \citenamefont {Peng}, \citenamefont {Sun},\ and\ \citenamefont
  {Perdew}}]{YangPRB16}%
  \BibitemOpen
  \bibfield  {author} {\bibinfo {author} {\bibfnamefont {Z.-h.}\ \bibnamefont
  {Yang}}, \bibinfo {author} {\bibfnamefont {H.}~\bibnamefont {Peng}}, \bibinfo
  {author} {\bibfnamefont {J.}~\bibnamefont {Sun}}, \ and\ \bibinfo {author}
  {\bibfnamefont {J.~P.}\ \bibnamefont {Perdew}},\ }\href@noop {} {\bibfield
  {journal} {\bibinfo  {journal} {Phys. Rev. B}\ }\textbf {\bibinfo {volume}
  {93}},\ \bibinfo {pages} {205205} (\bibinfo {year} {2016})}\BibitemShut
  {NoStop}%
\bibitem [{\citenamefont {Fuks}\ \emph {et~al.}(2011)\citenamefont {Fuks},
  \citenamefont {Rubio},\ and\ \citenamefont {Maitra}}]{FuksPRA11}%
  \BibitemOpen
  \bibfield  {author} {\bibinfo {author} {\bibfnamefont {J.~I.}\ \bibnamefont
  {Fuks}}, \bibinfo {author} {\bibfnamefont {A.}~\bibnamefont {Rubio}}, \ and\
  \bibinfo {author} {\bibfnamefont {N.~T.}\ \bibnamefont {Maitra}},\
  }\href@noop {} {\bibfield  {journal} {\bibinfo  {journal} {Phys. Rev. A}\
  }\textbf {\bibinfo {volume} {83}},\ \bibinfo {pages} {042501} (\bibinfo
  {year} {2011})}\BibitemShut {NoStop}%
\bibitem [{\citenamefont {Arnold}\ \emph {et~al.}(2011)\citenamefont {Arnold},
  \citenamefont {Siegmund},\ and\ \citenamefont {Pankratov}}]{ArnoldJPCM11}%
  \BibitemOpen
  \bibfield  {author} {\bibinfo {author} {\bibfnamefont {T.}~\bibnamefont
  {Arnold}}, \bibinfo {author} {\bibfnamefont {M.}~\bibnamefont {Siegmund}}, \
  and\ \bibinfo {author} {\bibfnamefont {O.}~\bibnamefont {Pankratov}},\
  }\href@noop {} {\bibfield  {journal} {\bibinfo  {journal} {J. Phys.: Condens.
  Matter}\ }\textbf {\bibinfo {volume} {23}},\ \bibinfo {pages} {335601}
  (\bibinfo {year} {2011})}\BibitemShut {NoStop}%
\bibitem [{\citenamefont {Qian}(2012)}]{QianPRB12}%
  \BibitemOpen
  \bibfield  {author} {\bibinfo {author} {\bibfnamefont {Z.}~\bibnamefont
  {Qian}},\ }\href@noop {} {\bibfield  {journal} {\bibinfo  {journal} {Phys.
  Rev. B}\ }\textbf {\bibinfo {volume} {85}},\ \bibinfo {pages} {115124}
  (\bibinfo {year} {2012})}\BibitemShut {NoStop}%
\bibitem [{\citenamefont {Vilhena}\ \emph {et~al.}(2012)\citenamefont
  {Vilhena}, \citenamefont {R\"as\"anen}, \citenamefont {Lehtovaara},\ and\
  \citenamefont {Marques}}]{VilhenaPRA12}%
  \BibitemOpen
  \bibfield  {author} {\bibinfo {author} {\bibfnamefont {J.~G.}\ \bibnamefont
  {Vilhena}}, \bibinfo {author} {\bibfnamefont {E.}~\bibnamefont
  {R\"as\"anen}}, \bibinfo {author} {\bibfnamefont {L.}~\bibnamefont
  {Lehtovaara}}, \ and\ \bibinfo {author} {\bibfnamefont {M.~A.~L.}\
  \bibnamefont {Marques}},\ }\href@noop {} {\bibfield  {journal} {\bibinfo
  {journal} {Phys. Rev. A}\ }\textbf {\bibinfo {volume} {85}},\ \bibinfo
  {pages} {052514} (\bibinfo {year} {2012})}\BibitemShut {NoStop}%
\bibitem [{\citenamefont {Schmidt}\ \emph {et~al.}(2014)\citenamefont
  {Schmidt}, \citenamefont {Kraisler}, \citenamefont {Kronik},\ and\
  \citenamefont {K\"{u}mmel}}]{SchmidtPCCP14}%
  \BibitemOpen
  \bibfield  {author} {\bibinfo {author} {\bibfnamefont {T.}~\bibnamefont
  {Schmidt}}, \bibinfo {author} {\bibfnamefont {E.}~\bibnamefont {Kraisler}},
  \bibinfo {author} {\bibfnamefont {L.}~\bibnamefont {Kronik}}, \ and\ \bibinfo
  {author} {\bibfnamefont {S.}~\bibnamefont {K\"{u}mmel}},\ }\href@noop {}
  {\bibfield  {journal} {\bibinfo  {journal} {Phys. Chem. Chem. Phys.}\
  }\textbf {\bibinfo {volume} {16}},\ \bibinfo {pages} {14357} (\bibinfo {year}
  {2014})}\BibitemShut {NoStop}%
\bibitem [{\citenamefont {Kraisler}\ \emph {et~al.}(2015)\citenamefont
  {Kraisler}, \citenamefont {Schmidt}, \citenamefont {K\"{u}mmel},\ and\
  \citenamefont {Kronik}}]{KraislerJCP15}%
  \BibitemOpen
  \bibfield  {author} {\bibinfo {author} {\bibfnamefont {E.}~\bibnamefont
  {Kraisler}}, \bibinfo {author} {\bibfnamefont {T.}~\bibnamefont {Schmidt}},
  \bibinfo {author} {\bibfnamefont {S.}~\bibnamefont {K\"{u}mmel}}, \ and\
  \bibinfo {author} {\bibfnamefont {L.}~\bibnamefont {Kronik}},\ }\href@noop {}
  {\bibfield  {journal} {\bibinfo  {journal} {J. Chem. Phys.}\ }\textbf
  {\bibinfo {volume} {143}},\ \bibinfo {pages} {104105} (\bibinfo {year}
  {2015})}\BibitemShut {NoStop}%
\bibitem [{\citenamefont {Kim}\ \emph {et~al.}(2015{\natexlab{a}})\citenamefont
  {Kim}, \citenamefont {Hong}, \citenamefont {Choi},\ and\ \citenamefont
  {Kim}}]{KimBKCS15}%
  \BibitemOpen
  \bibfield  {author} {\bibinfo {author} {\bibfnamefont {J.}~\bibnamefont
  {Kim}}, \bibinfo {author} {\bibfnamefont {K.}~\bibnamefont {Hong}}, \bibinfo
  {author} {\bibfnamefont {S.}~\bibnamefont {Choi}}, \ and\ \bibinfo {author}
  {\bibfnamefont {W.~Y.}\ \bibnamefont {Kim}},\ }\href@noop {} {\bibfield
  {journal} {\bibinfo  {journal} {Bull. Korean Chem. Soc.}\ }\textbf {\bibinfo
  {volume} {36}},\ \bibinfo {pages} {998} (\bibinfo {year}
  {2015}{\natexlab{a}})}\BibitemShut {NoStop}%
\bibitem [{\citenamefont {Kim}\ \emph {et~al.}(2015{\natexlab{b}})\citenamefont
  {Kim}, \citenamefont {Hong}, \citenamefont {Choi}, \citenamefont {Hwang},\
  and\ \citenamefont {Kim}}]{KimPCCP15}%
  \BibitemOpen
  \bibfield  {author} {\bibinfo {author} {\bibfnamefont {J.}~\bibnamefont
  {Kim}}, \bibinfo {author} {\bibfnamefont {K.}~\bibnamefont {Hong}}, \bibinfo
  {author} {\bibfnamefont {S.}~\bibnamefont {Choi}}, \bibinfo {author}
  {\bibfnamefont {S.-Y.}\ \bibnamefont {Hwang}}, \ and\ \bibinfo {author}
  {\bibfnamefont {W.~Y.}\ \bibnamefont {Kim}},\ }\href@noop {} {\bibfield
  {journal} {\bibinfo  {journal} {Phys. Chem. Chem. Phys.}\ }\textbf {\bibinfo
  {volume} {17}},\ \bibinfo {pages} {31434} (\bibinfo {year}
  {2015}{\natexlab{b}})}\BibitemShut {NoStop}%
\bibitem [{\citenamefont {Schmidt}\ and\ \citenamefont
  {K\"ummel}(2016)}]{SchmidtPRB16}%
  \BibitemOpen
  \bibfield  {author} {\bibinfo {author} {\bibfnamefont {T.}~\bibnamefont
  {Schmidt}}\ and\ \bibinfo {author} {\bibfnamefont {S.}~\bibnamefont
  {K\"ummel}},\ }\href@noop {} {\bibfield  {journal} {\bibinfo  {journal}
  {Phys. Rev. B}\ }\textbf {\bibinfo {volume} {93}},\ \bibinfo {pages} {165120}
  (\bibinfo {year} {2016})}\BibitemShut {NoStop}%
\bibitem [{\citenamefont {Gritsenko}\ \emph {et~al.}(1995)\citenamefont
  {Gritsenko}, \citenamefont {van Leeuwen}, \citenamefont {van Lenthe},\ and\
  \citenamefont {Baerends}}]{GritsenkoPRA95}%
  \BibitemOpen
  \bibfield  {author} {\bibinfo {author} {\bibfnamefont {O.}~\bibnamefont
  {Gritsenko}}, \bibinfo {author} {\bibfnamefont {R.}~\bibnamefont {van
  Leeuwen}}, \bibinfo {author} {\bibfnamefont {E.}~\bibnamefont {van Lenthe}},
  \ and\ \bibinfo {author} {\bibfnamefont {E.~J.}\ \bibnamefont {Baerends}},\
  }\href@noop {} {\bibfield  {journal} {\bibinfo  {journal} {Phys. Rev. A}\
  }\textbf {\bibinfo {volume} {51}},\ \bibinfo {pages} {1944} (\bibinfo {year}
  {1995})}\BibitemShut {NoStop}%
\bibitem [{\citenamefont {Grabo}\ and\ \citenamefont
  {Gross}(1995)}]{GraboCPL95}%
  \BibitemOpen
  \bibfield  {author} {\bibinfo {author} {\bibfnamefont {T.}~\bibnamefont
  {Grabo}}\ and\ \bibinfo {author} {\bibfnamefont {E.~K.~U.}\ \bibnamefont
  {Gross}},\ }\href@noop {} {\bibfield  {journal} {\bibinfo  {journal} {Chem.
  Phys. Lett.}\ }\textbf {\bibinfo {volume} {240}},\ \bibinfo {pages} {141}
  (\bibinfo {year} {1995})}\BibitemShut {NoStop}%
\bibitem [{\citenamefont {Grabo}\ \emph {et~al.}(1997)\citenamefont {Grabo},
  \citenamefont {Kreibich},\ and\ \citenamefont {Gross}}]{GraboME97}%
  \BibitemOpen
  \bibfield  {author} {\bibinfo {author} {\bibfnamefont {T.}~\bibnamefont
  {Grabo}}, \bibinfo {author} {\bibfnamefont {T.}~\bibnamefont {Kreibich}}, \
  and\ \bibinfo {author} {\bibfnamefont {E.~K.~U.}\ \bibnamefont {Gross}},\
  }\href@noop {} {\bibfield  {journal} {\bibinfo  {journal} {Mol. Eng.}\
  }\textbf {\bibinfo {volume} {7}},\ \bibinfo {pages} {27} (\bibinfo {year}
  {1997})}\BibitemShut {NoStop}%
\bibitem [{\citenamefont {Engel}\ \emph {et~al.}(2000)\citenamefont {Engel},
  \citenamefont {H\"{o}ck},\ and\ \citenamefont {Dreizler}}]{EngelPRA00}%
  \BibitemOpen
  \bibfield  {author} {\bibinfo {author} {\bibfnamefont {E.}~\bibnamefont
  {Engel}}, \bibinfo {author} {\bibfnamefont {A.}~\bibnamefont {H\"{o}ck}}, \
  and\ \bibinfo {author} {\bibfnamefont {R.~M.}\ \bibnamefont {Dreizler}},\
  }\href@noop {} {\bibfield  {journal} {\bibinfo  {journal} {Phys. Rev. A}\
  }\textbf {\bibinfo {volume} {62}},\ \bibinfo {pages} {042502} (\bibinfo
  {year} {2000})};\ \bibinfo {note} {\textbf{63}, 039901(E) (2001)}\BibitemShut
  {NoStop}%
\bibitem [{\citenamefont {Della~Sala}\ and\ \citenamefont
  {G\"{o}rling}(2001)}]{DellaSalaJCP01}%
  \BibitemOpen
  \bibfield  {author} {\bibinfo {author} {\bibfnamefont {F.}~\bibnamefont
  {Della~Sala}}\ and\ \bibinfo {author} {\bibfnamefont {A.}~\bibnamefont
  {G\"{o}rling}},\ }\href@noop {} {\bibfield  {journal} {\bibinfo  {journal}
  {J. Chem. Phys.}\ }\textbf {\bibinfo {volume} {115}},\ \bibinfo {pages}
  {5718} (\bibinfo {year} {2001})}\BibitemShut {NoStop}%
\bibitem [{\citenamefont {Gr\"{u}ning}\ \emph {et~al.}(2002)\citenamefont
  {Gr\"{u}ning}, \citenamefont {Gritsenko},\ and\ \citenamefont
  {Baerends}}]{GruningJCP02}%
  \BibitemOpen
  \bibfield  {author} {\bibinfo {author} {\bibfnamefont {M.}~\bibnamefont
  {Gr\"{u}ning}}, \bibinfo {author} {\bibfnamefont {O.~V.}\ \bibnamefont
  {Gritsenko}}, \ and\ \bibinfo {author} {\bibfnamefont {E.~J.}\ \bibnamefont
  {Baerends}},\ }\href@noop {} {\bibfield  {journal} {\bibinfo  {journal} {J.
  Chem. Phys.}\ }\textbf {\bibinfo {volume} {116}},\ \bibinfo {pages} {6435}
  (\bibinfo {year} {2002})}\BibitemShut {NoStop}%
\bibitem [{\citenamefont {K\"ummel}\ and\ \citenamefont
  {Perdew}(2003{\natexlab{a}})}]{KuemmelPRL03}%
  \BibitemOpen
  \bibfield  {author} {\bibinfo {author} {\bibfnamefont {S.}~\bibnamefont
  {K\"ummel}}\ and\ \bibinfo {author} {\bibfnamefont {J.~P.}\ \bibnamefont
  {Perdew}},\ }\href@noop {} {\bibfield  {journal} {\bibinfo  {journal} {Phys.
  Rev. Lett.}\ }\textbf {\bibinfo {volume} {90}},\ \bibinfo {pages} {043004}
  (\bibinfo {year} {2003}{\natexlab{a}})}\BibitemShut {NoStop}%
\bibitem [{\citenamefont {K\"ummel}\ and\ \citenamefont
  {Perdew}(2003{\natexlab{b}})}]{KuemmelPRB03}%
  \BibitemOpen
  \bibfield  {author} {\bibinfo {author} {\bibfnamefont {S.}~\bibnamefont
  {K\"ummel}}\ and\ \bibinfo {author} {\bibfnamefont {J.~P.}\ \bibnamefont
  {Perdew}},\ }\href@noop {} {\bibfield  {journal} {\bibinfo  {journal} {Phys.
  Rev. B}\ }\textbf {\bibinfo {volume} {68}},\ \bibinfo {pages} {035103}
  (\bibinfo {year} {2003}{\natexlab{b}})}\BibitemShut {NoStop}%
\bibitem [{\citenamefont {Makmal}\ \emph {et~al.}(2009)\citenamefont {Makmal},
  \citenamefont {Armiento}, \citenamefont {Engel}, \citenamefont {Kronik},\
  and\ \citenamefont {K\"{u}mmel}}]{MakmalPRB09}%
  \BibitemOpen
  \bibfield  {author} {\bibinfo {author} {\bibfnamefont {A.}~\bibnamefont
  {Makmal}}, \bibinfo {author} {\bibfnamefont {R.}~\bibnamefont {Armiento}},
  \bibinfo {author} {\bibfnamefont {E.}~\bibnamefont {Engel}}, \bibinfo
  {author} {\bibfnamefont {L.}~\bibnamefont {Kronik}}, \ and\ \bibinfo {author}
  {\bibfnamefont {S.}~\bibnamefont {K\"{u}mmel}},\ }\href@noop {} {\bibfield
  {journal} {\bibinfo  {journal} {Phys. Rev. B}\ }\textbf {\bibinfo {volume}
  {80}},\ \bibinfo {pages} {161204(R)} (\bibinfo {year} {2009})}\BibitemShut
  {NoStop}%
\bibitem [{\citenamefont {Ryabinkin}\ \emph {et~al.}(2013)\citenamefont
  {Ryabinkin}, \citenamefont {Kananenka},\ and\ \citenamefont
  {Staroverov}}]{RyabinkinPRL13}%
  \BibitemOpen
  \bibfield  {author} {\bibinfo {author} {\bibfnamefont {I.~G.}\ \bibnamefont
  {Ryabinkin}}, \bibinfo {author} {\bibfnamefont {A.~A.}\ \bibnamefont
  {Kananenka}}, \ and\ \bibinfo {author} {\bibfnamefont {V.~N.}\ \bibnamefont
  {Staroverov}},\ }\href@noop {} {\bibfield  {journal} {\bibinfo  {journal}
  {Phys. Rev. Lett.}\ }\textbf {\bibinfo {volume} {111}},\ \bibinfo {pages}
  {013001} (\bibinfo {year} {2013})}\BibitemShut {NoStop}%
\bibitem [{\citenamefont {Kohut}\ \emph {et~al.}(2014)\citenamefont {Kohut},
  \citenamefont {Ryabinkin},\ and\ \citenamefont {Staroverov}}]{KohutJCP14}%
  \BibitemOpen
  \bibfield  {author} {\bibinfo {author} {\bibfnamefont {S.~V.}\ \bibnamefont
  {Kohut}}, \bibinfo {author} {\bibfnamefont {I.~G.}\ \bibnamefont
  {Ryabinkin}}, \ and\ \bibinfo {author} {\bibfnamefont {V.~N.}\ \bibnamefont
  {Staroverov}},\ }\href@noop {} {\bibfield  {journal} {\bibinfo  {journal} {J.
  Chem. Phys.}\ }\textbf {\bibinfo {volume} {140}},\ \bibinfo {pages} {18A535}
  (\bibinfo {year} {2014})}\BibitemShut {NoStop}%
\bibitem [{\citenamefont {Engel}\ and\ \citenamefont
  {Schmid}(2009)}]{EngelPRL09}%
  \BibitemOpen
  \bibfield  {author} {\bibinfo {author} {\bibfnamefont {E.}~\bibnamefont
  {Engel}}\ and\ \bibinfo {author} {\bibfnamefont {R.~N.}\ \bibnamefont
  {Schmid}},\ }\href@noop {} {\bibfield  {journal} {\bibinfo  {journal} {Phys.
  Rev. Lett.}\ }\textbf {\bibinfo {volume} {103}},\ \bibinfo {pages} {036404}
  (\bibinfo {year} {2009})}\BibitemShut {NoStop}%
\bibitem [{\citenamefont {Rigamonti}\ \emph {et~al.}(2015)\citenamefont
  {Rigamonti}, \citenamefont {Horowitz},\ and\ \citenamefont
  {Proetto}}]{RigamontiPRB15}%
  \BibitemOpen
  \bibfield  {author} {\bibinfo {author} {\bibfnamefont {S.}~\bibnamefont
  {Rigamonti}}, \bibinfo {author} {\bibfnamefont {C.~M.}\ \bibnamefont
  {Horowitz}}, \ and\ \bibinfo {author} {\bibfnamefont {C.~R.}\ \bibnamefont
  {Proetto}},\ }\href@noop {} {\bibfield  {journal} {\bibinfo  {journal} {Phys.
  Rev. B}\ }\textbf {\bibinfo {volume} {92}},\ \bibinfo {pages} {235145}
  (\bibinfo {year} {2015})}\BibitemShut {NoStop}%
\bibitem [{\citenamefont {Slater}(1951)}]{SlaterPR51}%
  \BibitemOpen
  \bibfield  {author} {\bibinfo {author} {\bibfnamefont {J.~C.}\ \bibnamefont
  {Slater}},\ }\href@noop {} {\bibfield  {journal} {\bibinfo  {journal} {Phys.
  Rev.}\ }\textbf {\bibinfo {volume} {81}},\ \bibinfo {pages} {385} (\bibinfo
  {year} {1951})}\BibitemShut {NoStop}%
\bibitem [{\citenamefont {Bylander}\ and\ \citenamefont
  {Kleinman}(1995{\natexlab{a}})}]{BylanderPRL95}%
  \BibitemOpen
  \bibfield  {author} {\bibinfo {author} {\bibfnamefont {D.~M.}\ \bibnamefont
  {Bylander}}\ and\ \bibinfo {author} {\bibfnamefont {L.}~\bibnamefont
  {Kleinman}},\ }\href@noop {} {\bibfield  {journal} {\bibinfo  {journal}
  {Phys. Rev. Lett.}\ }\textbf {\bibinfo {volume} {74}},\ \bibinfo {pages}
  {3660} (\bibinfo {year} {1995}{\natexlab{a}})}\BibitemShut {NoStop}%
\bibitem [{\citenamefont {Bylander}\ and\ \citenamefont
  {Kleinman}(1995{\natexlab{b}})}]{BylanderPRB95}%
  \BibitemOpen
  \bibfield  {author} {\bibinfo {author} {\bibfnamefont {D.~M.}\ \bibnamefont
  {Bylander}}\ and\ \bibinfo {author} {\bibfnamefont {L.}~\bibnamefont
  {Kleinman}},\ }\href@noop {} {\bibfield  {journal} {\bibinfo  {journal}
  {Phys. Rev. B}\ }\textbf {\bibinfo {volume} {52}},\ \bibinfo {pages} {14566}
  (\bibinfo {year} {1995}{\natexlab{b}})}\BibitemShut {NoStop}%
\bibitem [{\citenamefont {Bylander}\ and\ \citenamefont
  {Kleinman}(1996)}]{BylanderPRB96}%
  \BibitemOpen
  \bibfield  {author} {\bibinfo {author} {\bibfnamefont {D.~M.}\ \bibnamefont
  {Bylander}}\ and\ \bibinfo {author} {\bibfnamefont {L.}~\bibnamefont
  {Kleinman}},\ }\href@noop {} {\bibfield  {journal} {\bibinfo  {journal}
  {Phys. Rev. B}\ }\textbf {\bibinfo {volume} {54}},\ \bibinfo {pages} {7891}
  (\bibinfo {year} {1996})}\BibitemShut {NoStop}%
\bibitem [{\citenamefont {Bylander}\ and\ \citenamefont
  {Kleinman}(1997)}]{BylanderPRB97}%
  \BibitemOpen
  \bibfield  {author} {\bibinfo {author} {\bibfnamefont {D.~M.}\ \bibnamefont
  {Bylander}}\ and\ \bibinfo {author} {\bibfnamefont {L.}~\bibnamefont
  {Kleinman}},\ }\href@noop {} {\bibfield  {journal} {\bibinfo  {journal}
  {Phys. Rev. B}\ }\textbf {\bibinfo {volume} {55}},\ \bibinfo {pages} {9432}
  (\bibinfo {year} {1997})}\BibitemShut {NoStop}%
\bibitem [{\citenamefont {Gritsenko}\ and\ \citenamefont
  {Baerends}(2001)}]{GritsenkoPRA01}%
  \BibitemOpen
  \bibfield  {author} {\bibinfo {author} {\bibfnamefont {O.~V.}\ \bibnamefont
  {Gritsenko}}\ and\ \bibinfo {author} {\bibfnamefont {E.~J.}\ \bibnamefont
  {Baerends}},\ }\href@noop {} {\bibfield  {journal} {\bibinfo  {journal}
  {Phys. Rev. A}\ }\textbf {\bibinfo {volume} {64}},\ \bibinfo {pages} {042506}
  (\bibinfo {year} {2001})}\BibitemShut {NoStop}%
\bibitem [{\citenamefont {Nagy}(1997)}]{NagyPRA97}%
  \BibitemOpen
  \bibfield  {author} {\bibinfo {author} {\bibfnamefont {{\'A}.}~\bibnamefont
  {Nagy}},\ }\href@noop {} {\bibfield  {journal} {\bibinfo  {journal} {Phys.
  Rev. A}\ }\textbf {\bibinfo {volume} {55}},\ \bibinfo {pages} {3465}
  (\bibinfo {year} {1997})}\BibitemShut {NoStop}%
\bibitem [{\citenamefont {Engel}\ \emph {et~al.}(2001)\citenamefont {Engel},
  \citenamefont {H\"{o}ck}, \citenamefont {Schmid}, \citenamefont {Dreizler},\
  and\ \citenamefont {Chetty}}]{EngelPRB01}%
  \BibitemOpen
  \bibfield  {author} {\bibinfo {author} {\bibfnamefont {E.}~\bibnamefont
  {Engel}}, \bibinfo {author} {\bibfnamefont {A.}~\bibnamefont {H\"{o}ck}},
  \bibinfo {author} {\bibfnamefont {R.~N.}\ \bibnamefont {Schmid}}, \bibinfo
  {author} {\bibfnamefont {R.~M.}\ \bibnamefont {Dreizler}}, \ and\ \bibinfo
  {author} {\bibfnamefont {N.}~\bibnamefont {Chetty}},\ }\href@noop {}
  {\bibfield  {journal} {\bibinfo  {journal} {Phys. Rev. B}\ }\textbf {\bibinfo
  {volume} {64}},\ \bibinfo {pages} {125111} (\bibinfo {year}
  {2001})}\BibitemShut {NoStop}%
\bibitem [{\citenamefont {Engel}(2014{\natexlab{b}})}]{EngelPRB14}%
  \BibitemOpen
  \bibfield  {author} {\bibinfo {author} {\bibfnamefont {E.}~\bibnamefont
  {Engel}},\ }\href@noop {} {\bibfield  {journal} {\bibinfo  {journal} {Phys.
  Rev. B}\ }\textbf {\bibinfo {volume} {89}},\ \bibinfo {pages} {245105}
  (\bibinfo {year} {2014}{\natexlab{b}})}\BibitemShut {NoStop}%
\bibitem [{\citenamefont {Natan}(2015)}]{NatanPCCP15}%
  \BibitemOpen
  \bibfield  {author} {\bibinfo {author} {\bibfnamefont {A.}~\bibnamefont
  {Natan}},\ }\href@noop {} {\bibfield  {journal} {\bibinfo  {journal} {Phys.
  Chem. Chem. Phys.}\ }\textbf {\bibinfo {volume} {17}},\ \bibinfo {pages}
  {31510} (\bibinfo {year} {2015})}\BibitemShut {NoStop}%
\bibitem [{\citenamefont {S\"{u}le}\ \emph {et~al.}(2000)\citenamefont
  {S\"{u}le}, \citenamefont {Kurth},\ and\ \citenamefont
  {Van~Doren}}]{SuleJCP00}%
  \BibitemOpen
  \bibfield  {author} {\bibinfo {author} {\bibfnamefont {P.}~\bibnamefont
  {S\"{u}le}}, \bibinfo {author} {\bibfnamefont {S.}~\bibnamefont {Kurth}}, \
  and\ \bibinfo {author} {\bibfnamefont {V.}~\bibnamefont {Van~Doren}},\
  }\href@noop {} {\bibfield  {journal} {\bibinfo  {journal} {J. Chem. Phys.}\
  }\textbf {\bibinfo {volume} {112}},\ \bibinfo {pages} {7355} (\bibinfo {year}
  {2000})}\BibitemShut {NoStop}%
\bibitem [{\citenamefont {Fukazawa}\ and\ \citenamefont
  {Akai}(2010)}]{FukazawaJPCM10}%
  \BibitemOpen
  \bibfield  {author} {\bibinfo {author} {\bibfnamefont {T.}~\bibnamefont
  {Fukazawa}}\ and\ \bibinfo {author} {\bibfnamefont {H.}~\bibnamefont
  {Akai}},\ }\href@noop {} {\bibfield  {journal} {\bibinfo  {journal} {J.
  Phys.: Condens. Matter}\ }\textbf {\bibinfo {volume} {22}},\ \bibinfo {pages}
  {405501} (\bibinfo {year} {2010})}\BibitemShut {NoStop}%
\bibitem [{\citenamefont {Fukazawa}\ and\ \citenamefont
  {Akai}(2015)}]{FukazawaJPCM15}%
  \BibitemOpen
  \bibfield  {author} {\bibinfo {author} {\bibfnamefont {T.}~\bibnamefont
  {Fukazawa}}\ and\ \bibinfo {author} {\bibfnamefont {H.}~\bibnamefont
  {Akai}},\ }\href@noop {} {\bibfield  {journal} {\bibinfo  {journal} {J.
  Phys.: Condens. Matter}\ }\textbf {\bibinfo {volume} {27}},\ \bibinfo {pages}
  {115502} (\bibinfo {year} {2015})}\BibitemShut {NoStop}%
\bibitem [{\citenamefont {Xu}\ and\ \citenamefont {Holzwarth}(2011)}]{XuPRB11}%
  \BibitemOpen
  \bibfield  {author} {\bibinfo {author} {\bibfnamefont {X.}~\bibnamefont
  {Xu}}\ and\ \bibinfo {author} {\bibfnamefont {N.~A.~W.}\ \bibnamefont
  {Holzwarth}},\ }\href@noop {} {\bibfield  {journal} {\bibinfo  {journal}
  {Phys. Rev. B}\ }\textbf {\bibinfo {volume} {84}},\ \bibinfo {pages} {155113}
  (\bibinfo {year} {2011})}\BibitemShut {NoStop}%
\bibitem [{\citenamefont {Blaha}\ \emph {et~al.}(2001)\citenamefont {Blaha},
  \citenamefont {Schwarz}, \citenamefont {Madsen}, \citenamefont {Kvasnicka},\
  and\ \citenamefont {Luitz}}]{WIEN2k}%
  \BibitemOpen
  \bibfield  {author} {\bibinfo {author} {\bibfnamefont {P.}~\bibnamefont
  {Blaha}}, \bibinfo {author} {\bibfnamefont {K.}~\bibnamefont {Schwarz}},
  \bibinfo {author} {\bibfnamefont {G.~K.~H.}\ \bibnamefont {Madsen}}, \bibinfo
  {author} {\bibfnamefont {D.}~\bibnamefont {Kvasnicka}}, \ and\ \bibinfo
  {author} {\bibfnamefont {J.}~\bibnamefont {Luitz}},\ }\href@noop {} {\emph
  {\bibinfo {title} {WIEN2K: An Augmented Plane Wave plus Local Orbitals
  Program for Calculating Crystal Properties}}}\ (\bibinfo  {publisher} {Vienna
  University of Technology},\ \bibinfo {address} {Austria},\ \bibinfo {year}
  {2001})\BibitemShut {NoStop}%
\bibitem [{\citenamefont {Andersen}(1975)}]{AndersenPRB75}%
  \BibitemOpen
  \bibfield  {author} {\bibinfo {author} {\bibfnamefont {O.~K.}\ \bibnamefont
  {Andersen}},\ }\href@noop {} {\bibfield  {journal} {\bibinfo  {journal}
  {Phys. Rev. B}\ }\textbf {\bibinfo {volume} {12}},\ \bibinfo {pages} {3060}
  (\bibinfo {year} {1975})}\BibitemShut {NoStop}%
\bibitem [{\citenamefont {Singh}\ and\ \citenamefont
  {Nordstr{\"{o}}m}(2006)}]{Singh}%
  \BibitemOpen
  \bibfield  {author} {\bibinfo {author} {\bibfnamefont {D.~J.}\ \bibnamefont
  {Singh}}\ and\ \bibinfo {author} {\bibfnamefont {L.}~\bibnamefont
  {Nordstr{\"{o}}m}},\ }\href@noop {} {\emph {\bibinfo {title} {Planewaves,
  Pseudopotentials and the LAPW Method, 2nd ed.}}}\ (\bibinfo  {publisher}
  {Springer},\ \bibinfo {address} {Berlin},\ \bibinfo {year}
  {2006})\BibitemShut {NoStop}%
\bibitem [{\citenamefont {Bl\"{u}gel}\ and\ \citenamefont
  {Bihlmayer}(2006)}]{Blugel}%
  \BibitemOpen
  \bibfield  {author} {\bibinfo {author} {\bibfnamefont {S.}~\bibnamefont
  {Bl\"{u}gel}}\ and\ \bibinfo {author} {\bibfnamefont {G.}~\bibnamefont
  {Bihlmayer}},\ }\href@noop {} {\emph {\bibinfo {title} {Computational
  Nanoscience: Do it Yourself!}}}\ (\bibinfo {address} {Forschungszentrum
  J\"{u}lich GmbH},\ \bibinfo {year} {2006})\ p.~\bibinfo {pages}
  {85}\BibitemShut {NoStop}%
\bibitem [{\citenamefont {Tran}\ \emph
  {et~al.}(2015{\natexlab{b}})\citenamefont {Tran}, \citenamefont {Blaha},\
  and\ \citenamefont {Schwarz}}]{TranJCTC15}%
  \BibitemOpen
  \bibfield  {author} {\bibinfo {author} {\bibfnamefont {F.}~\bibnamefont
  {Tran}}, \bibinfo {author} {\bibfnamefont {P.}~\bibnamefont {Blaha}}, \ and\
  \bibinfo {author} {\bibfnamefont {K.}~\bibnamefont {Schwarz}},\ }\href@noop
  {} {\bibfield  {journal} {\bibinfo  {journal} {J. Chem. Theory Comput.}\
  }\textbf {\bibinfo {volume} {11}},\ \bibinfo {pages} {4717} (\bibinfo {year}
  {2015}{\natexlab{b}})}\BibitemShut {NoStop}%
\bibitem [{SM_()}]{SM_KLI}%
  \BibitemOpen
  \href@noop {} {}\bibinfo {howpublished} {See Supplemental Material at
  http://link.aps.org/supplemental/ for information about the solids in the
  test set and the detailed equations of the KLI potential for the LAPW basis
  set.}\BibitemShut {Stop}%
\bibitem [{\citenamefont {Weinert}(1981)}]{WeinertJMP81}%
  \BibitemOpen
  \bibfield  {author} {\bibinfo {author} {\bibfnamefont {M.}~\bibnamefont
  {Weinert}},\ }\href@noop {} {\bibfield  {journal} {\bibinfo  {journal} {J.
  Math. Phys.}\ }\textbf {\bibinfo {volume} {22}},\ \bibinfo {pages} {2433}
  (\bibinfo {year} {1981})}\BibitemShut {NoStop}%
\bibitem [{\citenamefont {Massidda}\ \emph {et~al.}(1993)\citenamefont
  {Massidda}, \citenamefont {Posternak},\ and\ \citenamefont
  {Baldereschi}}]{MassiddaPRB93}%
  \BibitemOpen
  \bibfield  {author} {\bibinfo {author} {\bibfnamefont {S.}~\bibnamefont
  {Massidda}}, \bibinfo {author} {\bibfnamefont {M.}~\bibnamefont {Posternak}},
  \ and\ \bibinfo {author} {\bibfnamefont {A.}~\bibnamefont {Baldereschi}},\
  }\href@noop {} {\bibfield  {journal} {\bibinfo  {journal} {Phys. Rev. B}\
  }\textbf {\bibinfo {volume} {48}},\ \bibinfo {pages} {5058} (\bibinfo {year}
  {1993})}\BibitemShut {NoStop}%
\bibitem [{\citenamefont {Onida}\ \emph {et~al.}(1995)\citenamefont {Onida},
  \citenamefont {Reining}, \citenamefont {Godby}, \citenamefont {Del~Sole},\
  and\ \citenamefont {Andreoni}}]{OnidaPRL95}%
  \BibitemOpen
  \bibfield  {author} {\bibinfo {author} {\bibfnamefont {G.}~\bibnamefont
  {Onida}}, \bibinfo {author} {\bibfnamefont {L.}~\bibnamefont {Reining}},
  \bibinfo {author} {\bibfnamefont {R.~W.}\ \bibnamefont {Godby}}, \bibinfo
  {author} {\bibfnamefont {R.}~\bibnamefont {Del~Sole}}, \ and\ \bibinfo
  {author} {\bibfnamefont {W.}~\bibnamefont {Andreoni}},\ }\href@noop {}
  {\bibfield  {journal} {\bibinfo  {journal} {Phys. Rev. Lett.}\ }\textbf
  {\bibinfo {volume} {75}},\ \bibinfo {pages} {818} (\bibinfo {year}
  {1995})}\BibitemShut {NoStop}%
\bibitem [{\citenamefont {Spencer}\ and\ \citenamefont
  {Alavi}(2008)}]{SpencerPRB08}%
  \BibitemOpen
  \bibfield  {author} {\bibinfo {author} {\bibfnamefont {J.}~\bibnamefont
  {Spencer}}\ and\ \bibinfo {author} {\bibfnamefont {A.}~\bibnamefont
  {Alavi}},\ }\href@noop {} {\bibfield  {journal} {\bibinfo  {journal} {Phys.
  Rev. B}\ }\textbf {\bibinfo {volume} {77}},\ \bibinfo {pages} {193110}
  (\bibinfo {year} {2008})}\BibitemShut {NoStop}%
\bibitem [{\citenamefont {Tran}\ and\ \citenamefont {Blaha}(2011)}]{TranPRB11}%
  \BibitemOpen
  \bibfield  {author} {\bibinfo {author} {\bibfnamefont {F.}~\bibnamefont
  {Tran}}\ and\ \bibinfo {author} {\bibfnamefont {P.}~\bibnamefont {Blaha}},\
  }\href@noop {} {\bibfield  {journal} {\bibinfo  {journal} {Phys. Rev. B}\
  }\textbf {\bibinfo {volume} {83}},\ \bibinfo {pages} {235118} (\bibinfo
  {year} {2011})}\BibitemShut {NoStop}%
\bibitem [{\citenamefont {Engel}()}]{Engel16}%
  \BibitemOpen
  \bibfield  {author} {\bibinfo {author} {\bibfnamefont {E.}~\bibnamefont
  {Engel}},\ }\href@noop {} {}\bibinfo {howpublished} {(private
  communication)}\BibitemShut {NoStop}%
\bibitem [{\citenamefont {Betzinger}\ \emph {et~al.}(2010)\citenamefont
  {Betzinger}, \citenamefont {Friedrich},\ and\ \citenamefont
  {Bl\"ugel}}]{BetzingerPRB10}%
  \BibitemOpen
  \bibfield  {author} {\bibinfo {author} {\bibfnamefont {M.}~\bibnamefont
  {Betzinger}}, \bibinfo {author} {\bibfnamefont {C.}~\bibnamefont
  {Friedrich}}, \ and\ \bibinfo {author} {\bibfnamefont {S.}~\bibnamefont
  {Bl\"ugel}},\ }\href@noop {} {\bibfield  {journal} {\bibinfo  {journal}
  {Phys. Rev. B}\ }\textbf {\bibinfo {volume} {81}},\ \bibinfo {pages} {195117}
  (\bibinfo {year} {2010})}\BibitemShut {NoStop}%
\bibitem [{FLE()}]{FLEUR}%
  \BibitemOpen
  \href@noop {} {}\bibinfo {howpublished} {See http://www.flapw.de}\BibitemShut
  {NoStop}%
\bibitem [{\citenamefont {Betzinger}\ \emph {et~al.}(2012)\citenamefont
  {Betzinger}, \citenamefont {Friedrich}, \citenamefont {G\"orling},\ and\
  \citenamefont {Bl\"ugel}}]{BetzingerPRB12}%
  \BibitemOpen
  \bibfield  {author} {\bibinfo {author} {\bibfnamefont {M.}~\bibnamefont
  {Betzinger}}, \bibinfo {author} {\bibfnamefont {C.}~\bibnamefont
  {Friedrich}}, \bibinfo {author} {\bibfnamefont {A.}~\bibnamefont
  {G\"orling}}, \ and\ \bibinfo {author} {\bibfnamefont {S.}~\bibnamefont
  {Bl\"ugel}},\ }\href@noop {} {\bibfield  {journal} {\bibinfo  {journal}
  {Phys. Rev. B}\ }\textbf {\bibinfo {volume} {85}},\ \bibinfo {pages} {245124}
  (\bibinfo {year} {2012})}\BibitemShut {NoStop}%
\bibitem [{\citenamefont {Betzinger}\ \emph {et~al.}(2013)\citenamefont
  {Betzinger}, \citenamefont {Friedrich},\ and\ \citenamefont
  {Bl\"ugel}}]{BetzingerPRB13}%
  \BibitemOpen
  \bibfield  {author} {\bibinfo {author} {\bibfnamefont {M.}~\bibnamefont
  {Betzinger}}, \bibinfo {author} {\bibfnamefont {C.}~\bibnamefont
  {Friedrich}}, \ and\ \bibinfo {author} {\bibfnamefont {S.}~\bibnamefont
  {Bl\"ugel}},\ }\href@noop {} {\bibfield  {journal} {\bibinfo  {journal}
  {Phys. Rev. B}\ }\textbf {\bibinfo {volume} {88}},\ \bibinfo {pages} {075130}
  (\bibinfo {year} {2013})}\BibitemShut {NoStop}%
\bibitem [{\citenamefont {Friedrich}\ \emph {et~al.}(2013)\citenamefont
  {Friedrich}, \citenamefont {Betzinger},\ and\ \citenamefont
  {Bl\"ugel}}]{FriedrichPRA13}%
  \BibitemOpen
  \bibfield  {author} {\bibinfo {author} {\bibfnamefont {C.}~\bibnamefont
  {Friedrich}}, \bibinfo {author} {\bibfnamefont {M.}~\bibnamefont
  {Betzinger}}, \ and\ \bibinfo {author} {\bibfnamefont {S.}~\bibnamefont
  {Bl\"ugel}},\ }\href@noop {} {\bibfield  {journal} {\bibinfo  {journal}
  {Phys. Rev. A}\ }\textbf {\bibinfo {volume} {88}},\ \bibinfo {pages} {046501}
  (\bibinfo {year} {2013})}\BibitemShut {NoStop}%
\bibitem [{\citenamefont {Perdew}\ \emph {et~al.}(1996)\citenamefont {Perdew},
  \citenamefont {Burke},\ and\ \citenamefont {Ernzerhof}}]{PerdewPRL96}%
  \BibitemOpen
  \bibfield  {author} {\bibinfo {author} {\bibfnamefont {J.~P.}\ \bibnamefont
  {Perdew}}, \bibinfo {author} {\bibfnamefont {K.}~\bibnamefont {Burke}}, \
  and\ \bibinfo {author} {\bibfnamefont {M.}~\bibnamefont {Ernzerhof}},\
  }\href@noop {} {\bibfield  {journal} {\bibinfo  {journal} {Phys. Rev. Lett.}\
  }\textbf {\bibinfo {volume} {77}},\ \bibinfo {pages} {3865} (\bibinfo {year}
  {1996})};\ \bibinfo {note} {\textbf{78}, 1396(E) (1997)}\BibitemShut
  {NoStop}%
\bibitem [{\citenamefont {Engel}\ and\ \citenamefont
  {Vosko}(1993)}]{EngelPRB93}%
  \BibitemOpen
  \bibfield  {author} {\bibinfo {author} {\bibfnamefont {E.}~\bibnamefont
  {Engel}}\ and\ \bibinfo {author} {\bibfnamefont {S.~H.}\ \bibnamefont
  {Vosko}},\ }\href@noop {} {\bibfield  {journal} {\bibinfo  {journal} {Phys.
  Rev. B}\ }\textbf {\bibinfo {volume} {47}},\ \bibinfo {pages} {13164}
  (\bibinfo {year} {1993})}\BibitemShut {NoStop}%
\bibitem [{\citenamefont {Armiento}\ and\ \citenamefont
  {K\"{u}mmel}(2013)}]{ArmientoPRL13}%
  \BibitemOpen
  \bibfield  {author} {\bibinfo {author} {\bibfnamefont {R.}~\bibnamefont
  {Armiento}}\ and\ \bibinfo {author} {\bibfnamefont {S.}~\bibnamefont
  {K\"{u}mmel}},\ }\href@noop {} {\bibfield  {journal} {\bibinfo  {journal}
  {Phys. Rev. Lett.}\ }\textbf {\bibinfo {volume} {111}},\ \bibinfo {pages}
  {036402} (\bibinfo {year} {2013})}\BibitemShut {NoStop}%
\bibitem [{\citenamefont {R\"{a}s\"{a}nen}\ \emph {et~al.}(2010)\citenamefont
  {R\"{a}s\"{a}nen}, \citenamefont {Pittalis},\ and\ \citenamefont
  {Proetto}}]{RasanenJCP10}%
  \BibitemOpen
  \bibfield  {author} {\bibinfo {author} {\bibfnamefont {E.}~\bibnamefont
  {R\"{a}s\"{a}nen}}, \bibinfo {author} {\bibfnamefont {S.}~\bibnamefont
  {Pittalis}}, \ and\ \bibinfo {author} {\bibfnamefont {C.~R.}\ \bibnamefont
  {Proetto}},\ }\href@noop {} {\bibfield  {journal} {\bibinfo  {journal} {J.
  Chem. Phys.}\ }\textbf {\bibinfo {volume} {132}},\ \bibinfo {pages} {044112}
  (\bibinfo {year} {2010})}\BibitemShut {NoStop}%
\bibitem [{\citenamefont {Koelling}\ and\ \citenamefont
  {Harmon}(1977)}]{KoellingJPC77}%
  \BibitemOpen
  \bibfield  {author} {\bibinfo {author} {\bibfnamefont {D.~D.}\ \bibnamefont
  {Koelling}}\ and\ \bibinfo {author} {\bibfnamefont {B.~N.}\ \bibnamefont
  {Harmon}},\ }\href@noop {} {\bibfield  {journal} {\bibinfo  {journal} {J.
  Phys. C: Solid State Phys.}\ }\textbf {\bibinfo {volume} {10}},\ \bibinfo
  {pages} {3107} (\bibinfo {year} {1977})}\BibitemShut {NoStop}%
\bibitem [{\citenamefont {St\"adele}\ \emph {et~al.}(1997)\citenamefont
  {St\"adele}, \citenamefont {Majewski}, \citenamefont {Vogl},\ and\
  \citenamefont {G\"orling}}]{StaedelePRL97}%
  \BibitemOpen
  \bibfield  {author} {\bibinfo {author} {\bibfnamefont {M.}~\bibnamefont
  {St\"adele}}, \bibinfo {author} {\bibfnamefont {J.~A.}\ \bibnamefont
  {Majewski}}, \bibinfo {author} {\bibfnamefont {P.}~\bibnamefont {Vogl}}, \
  and\ \bibinfo {author} {\bibfnamefont {A.}~\bibnamefont {G\"orling}},\
  }\href@noop {} {\bibfield  {journal} {\bibinfo  {journal} {Phys. Rev. Lett.}\
  }\textbf {\bibinfo {volume} {79}},\ \bibinfo {pages} {2089} (\bibinfo {year}
  {1997})}\BibitemShut {NoStop}%
\bibitem [{\citenamefont {Hollins}\ \emph {et~al.}(2012)\citenamefont
  {Hollins}, \citenamefont {Clark}, \citenamefont {Refson},\ and\ \citenamefont
  {Gidopoulos}}]{HollinsPRB12}%
  \BibitemOpen
  \bibfield  {author} {\bibinfo {author} {\bibfnamefont {T.~W.}\ \bibnamefont
  {Hollins}}, \bibinfo {author} {\bibfnamefont {S.~J.}\ \bibnamefont {Clark}},
  \bibinfo {author} {\bibfnamefont {K.}~\bibnamefont {Refson}}, \ and\ \bibinfo
  {author} {\bibfnamefont {N.~I.}\ \bibnamefont {Gidopoulos}},\ }\href@noop {}
  {\bibfield  {journal} {\bibinfo  {journal} {Phys. Rev. B}\ }\textbf {\bibinfo
  {volume} {85}},\ \bibinfo {pages} {235126} (\bibinfo {year}
  {2012})}\BibitemShut {NoStop}%
\bibitem [{\citenamefont {Engel}(2016)}]{EngelIJQC16}%
  \BibitemOpen
  \bibfield  {author} {\bibinfo {author} {\bibfnamefont {E.}~\bibnamefont
  {Engel}},\ }\href@noop {} {\bibfield  {journal} {\bibinfo  {journal} {Int. J.
  Quantum Chem.}\ }\textbf {\bibinfo {volume} {116}},\ \bibinfo {pages} {867}
  (\bibinfo {year} {2016})}\BibitemShut {NoStop}%
\bibitem [{\citenamefont {Dufek}\ \emph {et~al.}(1994)\citenamefont {Dufek},
  \citenamefont {Blaha},\ and\ \citenamefont {Schwarz}}]{DufekPRB94}%
  \BibitemOpen
  \bibfield  {author} {\bibinfo {author} {\bibfnamefont {P.}~\bibnamefont
  {Dufek}}, \bibinfo {author} {\bibfnamefont {P.}~\bibnamefont {Blaha}}, \ and\
  \bibinfo {author} {\bibfnamefont {K.}~\bibnamefont {Schwarz}},\ }\href@noop
  {} {\bibfield  {journal} {\bibinfo  {journal} {Phys. Rev. B}\ }\textbf
  {\bibinfo {volume} {50}},\ \bibinfo {pages} {7279} (\bibinfo {year}
  {1994})}\BibitemShut {NoStop}%
\bibitem [{\citenamefont {Tran}\ \emph {et~al.}(2007)\citenamefont {Tran},
  \citenamefont {Blaha},\ and\ \citenamefont {Schwarz}}]{TranJPCM07}%
  \BibitemOpen
  \bibfield  {author} {\bibinfo {author} {\bibfnamefont {F.}~\bibnamefont
  {Tran}}, \bibinfo {author} {\bibfnamefont {P.}~\bibnamefont {Blaha}}, \ and\
  \bibinfo {author} {\bibfnamefont {K.}~\bibnamefont {Schwarz}},\ }\href@noop
  {} {\bibfield  {journal} {\bibinfo  {journal} {J. Phys.: Condens. Matter}\
  }\textbf {\bibinfo {volume} {19}},\ \bibinfo {pages} {196208} (\bibinfo
  {year} {2007})}\BibitemShut {NoStop}%
\bibitem [{\citenamefont {Grabo}\ \emph {et~al.}(2000)\citenamefont {Grabo},
  \citenamefont {Kreibich}, \citenamefont {Kurth},\ and\ \citenamefont
  {Gross}}]{Grabo00}%
  \BibitemOpen
  \bibfield  {author} {\bibinfo {author} {\bibfnamefont {T.}~\bibnamefont
  {Grabo}}, \bibinfo {author} {\bibfnamefont {T.}~\bibnamefont {Kreibich}},
  \bibinfo {author} {\bibfnamefont {S.}~\bibnamefont {Kurth}}, \ and\ \bibinfo
  {author} {\bibfnamefont {E.~K.~U.}\ \bibnamefont {Gross}},\ }\href@noop {}
  {\emph {\bibinfo {title} {Strong Coulomb Correlations in Electronic Structure
  Calculations: Beyond the Local Density Approximation}}},\ edited by\ \bibinfo
  {editor} {\bibfnamefont {V.~I.}\ \bibnamefont {Anisimov}}\ (\bibinfo
  {publisher} {Gordon and Breach},\ \bibinfo {address} {New York},\ \bibinfo
  {year} {2000})\ pp.\ \bibinfo {pages} {203--311}\BibitemShut {NoStop}%
\bibitem [{\citenamefont {Koller}\ \emph {et~al.}(2011)\citenamefont {Koller},
  \citenamefont {Tran},\ and\ \citenamefont {Blaha}}]{KollerPRB11}%
  \BibitemOpen
  \bibfield  {author} {\bibinfo {author} {\bibfnamefont {D.}~\bibnamefont
  {Koller}}, \bibinfo {author} {\bibfnamefont {F.}~\bibnamefont {Tran}}, \ and\
  \bibinfo {author} {\bibfnamefont {P.}~\bibnamefont {Blaha}},\ }\href@noop {}
  {\bibfield  {journal} {\bibinfo  {journal} {Phys. Rev. B}\ }\textbf {\bibinfo
  {volume} {83}},\ \bibinfo {pages} {195134} (\bibinfo {year}
  {2011})}\BibitemShut {NoStop}%
\bibitem [{\citenamefont {St\"adele}\ \emph {et~al.}(1999)\citenamefont
  {St\"adele}, \citenamefont {Moukara}, \citenamefont {Majewski}, \citenamefont
  {Vogl},\ and\ \citenamefont {G\"orling}}]{StaedelePRB99}%
  \BibitemOpen
  \bibfield  {author} {\bibinfo {author} {\bibfnamefont {M.}~\bibnamefont
  {St\"adele}}, \bibinfo {author} {\bibfnamefont {M.}~\bibnamefont {Moukara}},
  \bibinfo {author} {\bibfnamefont {J.~A.}\ \bibnamefont {Majewski}}, \bibinfo
  {author} {\bibfnamefont {P.}~\bibnamefont {Vogl}}, \ and\ \bibinfo {author}
  {\bibfnamefont {A.}~\bibnamefont {G\"orling}},\ }\href@noop {} {\bibfield
  {journal} {\bibinfo  {journal} {Phys. Rev. B}\ }\textbf {\bibinfo {volume}
  {59}},\ \bibinfo {pages} {10031} (\bibinfo {year} {1999})}\BibitemShut
  {NoStop}%
\bibitem [{\citenamefont {Aulbur}\ \emph {et~al.}(2000)\citenamefont {Aulbur},
  \citenamefont {St\"adele},\ and\ \citenamefont {G\"orling}}]{AulburPRB00}%
  \BibitemOpen
  \bibfield  {author} {\bibinfo {author} {\bibfnamefont {W.~G.}\ \bibnamefont
  {Aulbur}}, \bibinfo {author} {\bibfnamefont {M.}~\bibnamefont {St\"adele}}, \
  and\ \bibinfo {author} {\bibfnamefont {A.}~\bibnamefont {G\"orling}},\
  }\href@noop {} {\bibfield  {journal} {\bibinfo  {journal} {Phys. Rev. B}\
  }\textbf {\bibinfo {volume} {62}},\ \bibinfo {pages} {7121} (\bibinfo {year}
  {2000})}\BibitemShut {NoStop}%
\bibitem [{\citenamefont {Qteish}\ \emph {et~al.}(2006)\citenamefont {Qteish},
  \citenamefont {Rinke}, \citenamefont {Scheffler},\ and\ \citenamefont
  {Neugebauer}}]{QteishPRB06}%
  \BibitemOpen
  \bibfield  {author} {\bibinfo {author} {\bibfnamefont {A.}~\bibnamefont
  {Qteish}}, \bibinfo {author} {\bibfnamefont {P.}~\bibnamefont {Rinke}},
  \bibinfo {author} {\bibfnamefont {M.}~\bibnamefont {Scheffler}}, \ and\
  \bibinfo {author} {\bibfnamefont {J.}~\bibnamefont {Neugebauer}},\
  }\href@noop {} {\bibfield  {journal} {\bibinfo  {journal} {Phys. Rev. B}\
  }\textbf {\bibinfo {volume} {74}},\ \bibinfo {pages} {245208} (\bibinfo
  {year} {2006})}\BibitemShut {NoStop}%
\bibitem [{\citenamefont {Becke}(1988)}]{BeckePRA88}%
  \BibitemOpen
  \bibfield  {author} {\bibinfo {author} {\bibfnamefont {A.~D.}\ \bibnamefont
  {Becke}},\ }\href@noop {} {\bibfield  {journal} {\bibinfo  {journal} {Phys.
  Rev. A}\ }\textbf {\bibinfo {volume} {38}},\ \bibinfo {pages} {3098}
  (\bibinfo {year} {1988})}\BibitemShut {NoStop}%
\end{thebibliography}%

\end{document}